\newcommand{\ket}[1]{\ensuremath{\left|  #1 \right\rangle}}
\newcommand{\bra}[1]{\ensuremath{\left\langle  #1 \right|}}

\newcommand{\SU}[0]{\ensuremath{\left|  \uparrow \right\rangle}}
\newcommand{\SD}[0]{\ensuremath{\left|  \downarrow \right\rangle}}

\newcommand{\ind}[1]{_{\text{#1}}}
\newcommand{\tr}{\text{tr}}

\documentclass[aps,reprint,superscriptaddress,pra]{revtex4-1}
\usepackage{amsmath,amssymb,graphicx,color,verbatim,soul}

\begin{document}

\title{Measuring out-of-time-order correlations and multiple quantum spectra in a trapped ion quantum magnet}

\author{Martin G\"arttner}
\thanks{These authors contributed equally.}
\affiliation{JILA, NIST and University of Colorado, Boulder, Colorado 80309, USA}
\author{Justin G.\ Bohnet}
\thanks{These authors contributed equally.}
\affiliation{NIST, Boulder, Colorado 80305, USA}
\author{Arghavan Safavi-Naini}
\affiliation{JILA, NIST and University of Colorado, Boulder, Colorado 80309, USA}
\author{Michael L.\ Wall}
\affiliation{JILA, NIST and University of Colorado, Boulder, Colorado 80309, USA}
\author{John J.\ Bollinger} \email{john.bollinger@nist.gov}
\affiliation{NIST, Boulder, Colorado 80305, USA}
\author{Ana Maria Rey} \email{arey@jila.colorado.edu}
\affiliation{JILA, NIST and Department of Physics, University of Colorado, Boulder, Colorado, 80309, USA}

\date{\today}

\begin{abstract}
Controllable arrays of ions and ultra-cold atoms can simulate complex many-body phenomena and may provide insights into unsolved problems in modern science. 
To this end, experimentally feasible protocols for quantifying the buildup of quantum correlations and coherence are needed, as performing full state tomography does not scale favorably with the number of particles.
Here we develop and experimentally demonstrate such a protocol, which uses time reversal of the many-body dynamics to measure out-of-time-order correlation functions (OTOCs) in a long-range Ising spin quantum simulator with more than 100 ions in a Penning trap. 
By measuring a family of OTOCs as a function of a tunable parameter we obtain fine-grained information about the state of the system encoded in the multiple quantum coherence spectrum, extract the quantum state purity, and demonstrate the buildup of up to 8-body correlations. 
Future applications of this protocol could enable studies of many-body localization, quantum phase transitions, and tests of the holographic duality between quantum and gravitational systems.
\end{abstract}

\maketitle

Time-reversal has fascinated and puzzled physicists for centuries. In an iconic example, Josef Loschmidt argued that the second law of thermodynamics would be violated by time-reversing an entropy-increasing collision \cite{Loschmidt1876}. Ludwig Boltzmann responded by formulating the probabilistic definition of entropy, one of the cornerstones of statistical mechanics, and,  now a fundamental concept in  quantum information. Since the days of Boltzmann and Loschmidt, the notion of time-reversal has moved from the arena of thought experiments into the laboratory, with time-reversal of non-interacting quantum systems in the form of Hahn spin echoes \cite{Hahn1950} forming an essential part of nuclear magnetic resonance (NMR) \cite{Baum1985} and magnetic resonance imaging.

Recently, the experimental implementation of \emph{many-body} time-reversal protocols \cite{Widera2008, Linnemann2016} in atomic quantum systems have attracted attention \cite{Swingle2016, Yao2016, Shen2016, Zhu2016} for their potential to quantify the flow of quantum information in time and set bounds on thermalization times \cite{Sekino2008, Shenker2014, Shenker2015, Eisert2015}, which might also enable experimental tests of the holographic duality between quantum and gravitational systems \cite{Wenbo2016, Danshita2016, Hosur2016, Kitaev2015, Swingle2016}.
The key quantities sought after are special types of out-of-time-order correlation (OTOC) functions,
\begin{equation}
    F(\tau)= \langle \hat{W}^\dagger(\tau) \hat{V}^\dagger \hat{W}(\tau) \hat{V}\rangle,
\label{eq:OTCF}
\end{equation}
where  $\hat{W}(\tau)=e^{i \hat{H} \tau}\hat{W} e^{-i \hat{H} \tau}$, with $\hat{H}$ an interacting many-body Hamiltonian  and $\hat{W}$ and $\hat{V}$ two commuting unitary operators. 
Physically, $F(\tau)$ measures the ``scrambling'' of quantum information across the system's many-body degrees of freedom, for example, how fast an initial local perturbation becomes inaccessible to local probes \cite{Hosur2016}. 
Since ${\rm Re}[F(\tau)]=1-\langle|[ \hat{W}(\tau), \hat{V}]|^2\rangle/2$, $ F(\tau)$ encapsulates the degree by  which the initially commuting operators $\hat{W}$ and $\hat{V}$ fail to commute at later times due to the interactions generated by $\hat{H}$, which we adopt as an operational definition of scrambling.

Most theoretical studies of scrambling have focused on so-called fast scramblers in thermal states \cite{Sekino2008,Shenker2014,Hosur2016}, systems where the commutator grows exponentially at a rate exclusively determined by the temperature. 
However, the scrambling behavior of non-equilibrium systems at zero temperature will depend on the microscopic parameters of the Hamiltonian. 
This largely unexplored topic can provide valuable insights into the dynamics of interacting quantum many-body systems.

Here we perform measurements of OTOCs with a quantum simulator composed of more than 100 trapped ions \cite{Bohnet2016} interacting via all-to-all Ising interactions that can be reversed in time. This Ising interaction allows us to study interesting entangled states \cite{Leibfried2005,Monz2011,Bohnet2016,Strobel2014}, yet still operate in a regime where simulations on conventional computers are feasible. Thus our work is a first stepping stone for exploring scrambling in initially pure quantum systems. 
Our approach is modeled after the multiple quantum coherence (MQC) protocol developed in the context of NMR \cite{Baum1985, Alvarez2015, Sanchez2009} to quantify the buildup of multi-particle coherences (off-diagonal elements of the many-body density matrix). 
We show that this protocol, under specific choices of the initial state (pure states in our experiment), implements the measurement of a family of OTOCs.
Careful comparison with theory allows us to use the measurements as a verification protocol to benchmark the performance of the quantum simulator and to sensitively quantify different sources of decoherence and imperfect control. 
In our experiment, which starts with a pure product state, scrambling can be physically interpreted as the process by which the information stored (or encoded) in the initial state, through the interactions, is distributed over and therefore “stored in” other many-body degrees of freedom of the system. Thus, it cannot be extracted by measurements of single particle observables. Instead it requires measurements of higher order correlations. The information is not lost, but requires reading out the various degrees of freedom.

Future generalizations such as adding a spatially inhomogeneous magnetic field or a periodic drive would allow to experimentally study scrambling behavior in regimes intractable to theory, to explore the possibility of fast scrambling in low temperature systems, and to investigate possible connections between chaos and fast scrambling away from the semi-classical limit.
The protocols demonstrated here are widely applicable and could be implemented in a variety of other platforms with reversible dynamics, such as linear ion chains \cite{Monz2011, Debnath2016}, ultracold atomic gases \cite{Widera2008, Linnemann2016, Cucchietti2010}, cold atoms in optical cavities \cite{Leroux2010, Hosten2016, Douglas2015}, Rydberg-dressed atoms \cite{Macri2016}, superconducting qubits \cite{Houck2012}, and NMR systems \cite{Alvarez2015}.

\begin{figure*}
    \centering
    \includegraphics[width=\textwidth]{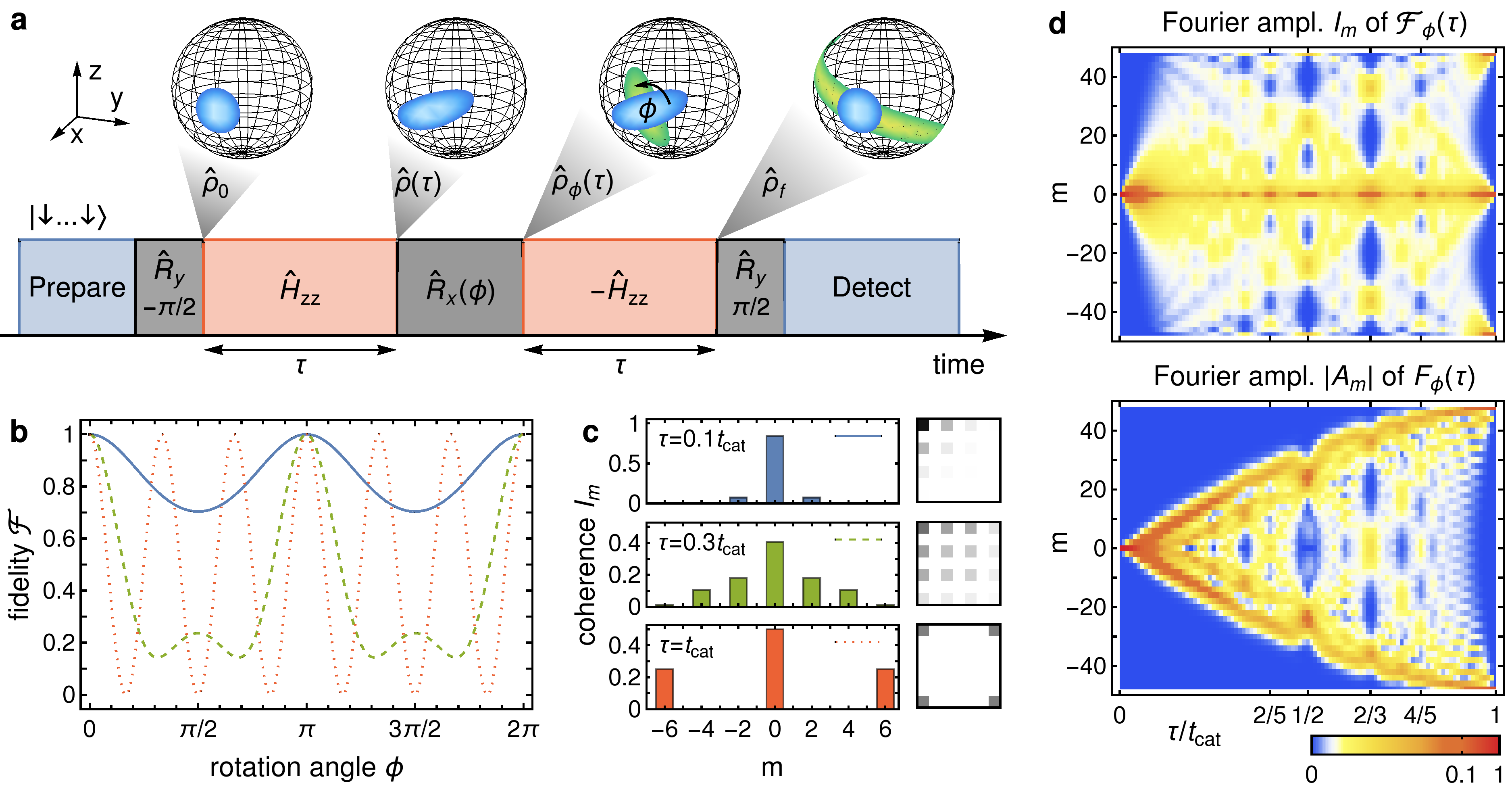}
    \caption{
    \textbf{Illustration of the many-body echo scheme.} \textbf{a}, Experimental sequence. The global $-\pi/2$ rotation $\hat{R}_y$ about the $y$-axis prepares an initial state with all spins pointing along the $x$-axis, and enables a measurement in this same basis. The generalized Bloch spheres illustrate the evolution of the state (Husimi distribution). In the case of $\phi=0$ (blue) the spins return to the initial state, while for $\phi=\pi/2$ (green) the overlap of the final state $\hat\rho_f$ with the initial state is small.
    \textbf{b}, Fidelity signal for an idealized case with $N=6$ spins and different evolution times $\tau$ given in \textbf{c}.
    \textbf{c}, The Fourier transforms of the fidelity signals of \textbf{b}. The Fourier amplitudes are identical to the MQCs $I_m$, which quantify the coherence of the state $\hat\rho(\tau)$. The small squares on the right show the absolute values of the density matrix elements of $\hat\rho(\tau)$ in the basis of symmetric Dicke states. Thus, $I_m$ is the sum of the squares of all matrix elements at a distance $m$ from the diagonal. The times are given in units of the time to reach the Schr\"odinger cat state $t\ind{cat}=\pi\hbar N/(4J)$. \textbf{d}, Simulated dynamics of the Fourier amplitudes of fidelity, $I_m$, and magnetization, $A_m$, for purely coherent evolution of $48$ ions, illustrating complementary probes of the flow of quantum information. The vanishing odd Fourier components are not shown.}
    \label{fig:seq}
\end{figure*}

The general protocol is illustrated in Fig.~\ref{fig:seq}a.
For concreteness, we consider the system of spin-$1/2$s, which we implement in our trapped ion experiment.
The state of interest $\hat\rho(\tau)$ is prepared by evolving a fiducial state, $\hat\rho_0$, under an interacting Hamiltonian $\hat H$ for a time $\tau$. In our experiment the initial density matrix is $\hat\rho_0=\ket{+\ldots+}\bra{+\ldots+}$, where $\ket{+}=(\ket{\uparrow}+\ket{\downarrow})/\sqrt{2}$ and $\hat H$ is a collective Ising model given by
\begin{equation}
\label{eq:Hspin}
    \hat H\ind{zz}=\frac{J}{N}\sum_{i<j} \hat \sigma_i^z \hat \sigma_j^z\, ,
\end{equation}where $N$ is the number of spins and $\hat \sigma_i^z$ are Pauli spin operators.
Inverting the sign of $\hat H$ (by changing $J$ to $-J$) and evolving again for time $\tau$ to the final state $\hat\rho_f$, implements the many-body time-reversal, which ideally takes the system back to the initial state $\hat\rho_0$.
If a state rotation $\hat{R}_x(\phi)=e^{-i \hat{S}_x\phi}$, here about the $x$-axis with $\hat{S}_x=\frac{1}{2}\sum_i\hat{\sigma}_i^x$, is inserted between the two halves of the time evolution through a variable angle $\phi$, the dependence of the revival probability on this angle contains information about $\hat\rho(\tau)$. In this work, we measure two different observables at the end of the sequence, the collective magnetization along the $x$-direction, $ \frac{2}{N}\langle \hat{S}_x\rangle=\frac{2}{N}\tr[\hat{S}_x\hat\rho_f]$, and the fidelity $\mathcal{F}_\phi(\tau)=\tr[\hat\rho_0 \hat\rho_f]$. 

The magnetization provides a direct measurement of
\begin{equation}
\label{eq:magne}
 \frac{2}{N}\langle \hat{S}_x\rangle =F_\phi(\tau)=\langle \hat{W}^\dagger_\phi(\tau) \hat{\sigma}^x_i \hat{W}_\phi(\tau) \hat{\sigma}^x_i\rangle_0,
\end{equation}
for any $i$, with $\hat{W}_\phi(\tau)=e^{i \hat{H}_{zz} \tau}\hat{R}_x(\phi)e^{-i \hat{H}_{zz} \tau}$. Here, $\langle\cdot\rangle_0$ denotes the expectation value in state $\hat\rho_0$. The implementation is facilitated by the fact that $\hat{V}\ket{+}=\hat{\sigma}_i^x\ket{+}=\ket{+}$. Moreover, single spin resolution is not necessary due to the permutation symmetry of our system that directly maps $\hat{\sigma}_i^x$ to the global magnetization along $x$: $\hat{\sigma}_i^x \to (2/N) \hat{S}_x$. In the absence of permutation symmetry, the OTOC measured by $F_\phi(\tau)$ should be interpreted as the average over the magnetization of each of the spins in the array.

Similarly, the fidelity, i.e.\ many-body overlap with the initial state, can be cast as an OTOC, where now $\hat{V}=\hat{\rho}_0$ is not unitary but $\mathcal{F}_\phi(\tau)$ still measures the failure of two operators to commute following dynamical evolution (see Methods).
Moreover, the fidelity can be directly linked to the so-called multiple quantum intensities $I_m$ (see Methods), which quantify the amplitudes of the off-diagonal elements \cite{Baum1985}, or coherences, of the density matrix $\hat\rho(\tau)$. The $I_m$ are measured by the Fourier components of
\begin{equation}
    \mathcal{F}_\phi(\tau) = \tr[\hat\rho_f\hat\rho_0] = \tr[\hat\rho(\tau)\hat\rho_\phi(\tau)]=\sum_{m=-N}^{N} I_m(\tau) e^{-i m\phi}\, ,
\label{eq:FtoMQC}
\end{equation}
where $\hat\rho_\phi(\tau)=\hat{R}_x(\phi)\hat\rho(\tau)\hat{R}_x^\dagger(\phi)$ (see Methods). 
In contrast to previous implementations in NMR spectroscopy, which typically operate at effectively infinite temperature, here we consider a spin system that is initially in a pure state at zero temperature.

Beyond the expected decay of the measured OTOCs for increasing $\tau$ and fixed $\phi$, studying the dependence of them on the rotation angle $\phi$ thus reveals information about the buildup of correlations and provides fine-grained information about the many-body properties of the state $\hat{\rho}(\tau)$.
The value of the fidelity at $\phi=0$ mod $2\pi$ also provides a direct measurement of the purity of the many-body spin state, $\mathcal{F}_0(\tau)=\tr[\hat\rho(\tau)^2]$. 
Note that the fidelity measurement directly implements a many-body Loschmidt echo, which is typically challenging to experimentally measure for systems of more than $\sim\! 10$ particles.

To clearly illustrate the dynamics of $I_m$ and their connection to off-diagonal elements of the density matrix, we first compute $\mathcal{F}_\phi(\tau)$ for a small system with  $N=6$ spins shown in Fig.~\ref{fig:seq}b and~\ref{fig:seq}c. At $\tau=\pi\hbar N/(4J)$ a macroscopic superposition (Schr\"odinger cat) state along $x$ is formed \cite{Sorensen1999}, which is signaled in the MQC spectrum by the cancellation of all $I_m$ except $I_0$ and $I_{\pm N}$. Note that for this case our scheme is equivalent to the interferometric cat-state verification scheme realized with $N\leq 6$ ions in Paul traps \cite{Leibfried2005}.

Motivated by the MQC protocol we study the dynamics of the Fourier amplitudes $A_m$ of the magnetization 
\begin{equation}
    F_\phi(\tau) = \sum_{m=-N}^{N} A_m(\tau) e^{-i m\phi}\, ,
\end{equation}
which probe the buildup of many-body correlations.
One can show that a non-zero $A_{m}(\tau)$ signals the buildup of at least $m$-body correlations. 
In the case of the Ising model, where all terms in the Hamiltonian commute with each other, $A_{m}(\tau)$ can only be non-zero if the Hamiltonian directly couples a given spin to $m-1$ other spins (see Methods and Supplementary Information).
In Fig.~\ref{fig:seq}d we illustrate the $I_m$ and $A_m$ dynamics for $N=48$ in the absence of decoherence, showing the sequential buildup of higher order coherences and correlations. Even for the homogeneous Ising interaction, the protocol reveals a rich structure in the many-body state, including multiple revivals of coherences. The $I_m$ spread more rapidly than the $A_m$ because the $I_m$ depends on the many-body overlap with the initial state, an $N$-body operator, which is more sensitive to the central rotation than the mean spin, a single-body observable.

\begin{figure}
    \centering
    \includegraphics[width=0.5\textwidth]{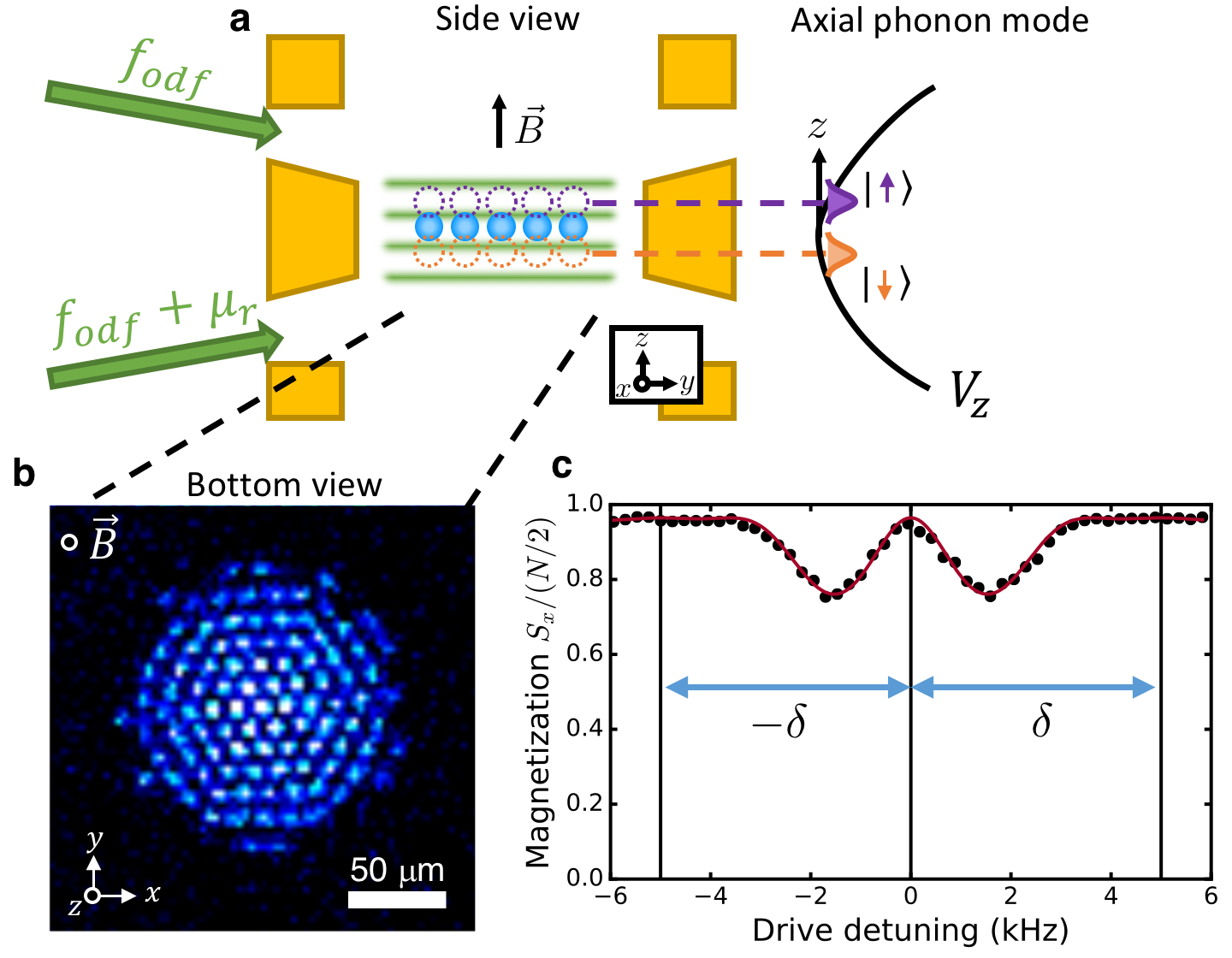}
    \caption{\textbf{Phonon-mediated, reversible spin-spin coupling in a Penning trap.}
    \textbf{a}, (left) Illustration of Penning trap cross-section. Ions (blue circles) are confined axially to a single 2D plane (shown in \textbf{b}) with static electric fields from potentials on the electrodes (gold). Rotation of the ions in the axial magnetic field $\vec{B}$ produces radial confinement from the Lorentz force. A pair of detuned ODF beams (green) interfere and form a traveling wave optical lattice, producing spin-dependent COM mode excitations that couple the spins to the axial phonon mode. Shown here are two of $(2N+1)$ excitations: all ions in $\ket{\uparrow}$ (purple) and all in $\ket{\downarrow}$ (orange). (right) The phonon wave packets experience equal and opposite displacement in the axial potential $V_z$. Spin-dependent motion, along with the Coulomb interaction, generates the spin-spin coupling.
    \textbf{b}, Rotating frame image of 2D array of $^9$Be$^+$ ions, integration time 2.1\,s.
    \textbf{c}, Residual spin-phonon coupling for drive frequencies away from the decoupling points $\pm \delta$ appears as a decrease in the magnetization measured after the experimental sequence from Fig.~\ref{fig:seq}, with  $\phi = \pi$, and without inverting $\hat H\ind{zz}$. Here $\tau = 200\,\mu$s. Note that decoupling points appear at $\pm \delta$ with $+\delta$ giving an anti-ferromagnetic interaction, and $-\delta$ giving a ferromagnetic interaction used for the time reversal of the $\hat H\ind{zz}$ dynamics.}
    \label{fig:expt}
\end{figure}

Our experimental demonstration uses 2D arrays of laser-cooled $^9\text{Be}^+$ ions in a Penning trap, where the spins are the valence electron spin states in the $B = 4.46$ T magnetic field \cite{Britton2012, Sawyer2012, Bohnet2016}.
Arbitrary collective spin rotations are applied via microwave pulses (see Fig.~\ref{fig:expt} and Supplementary Information).
Long-range, tunable spin interactions are engineered through a time-dependent optical dipole force (ODF), characterized by a frequency $\mu_r$, that couples the spins to the axial motional (phonon) modes of the ion crystal.
The driven spin-dependent motion, combined with the Coulomb force, mediates the spin-spin interaction.
Laser cooling and optical pumping allow us to initialize the spins in a pure, coherent collective spin state with fidelity $> 99.9\%$ \cite{Biercuk2009}, and initialize the motional modes with an average thermal occupation of $6$ quanta, set by the Doppler cooling limit.

To implement the reversible Ising dynamics, we operate in a regime where the spins couple to a single phonon mode, the collective center-of-mass (COM) mode at frequency $\omega_z$.
Although there are $N$ axial phonon modes in the crystal, the COM mode is well-resolved for the ODF detuning from the COM mode $\delta = \mu_r-\omega_z$ used here \cite{Sawyer2012}, justifying the single mode approximation.
Then the spin-phonon dynamics are given by \cite{Sorensen1999, Leibfried2003}
\begin{align}
\label{eq:HSP}
\hat{H}_{I}&=-\frac{\Omega_0}{2\sqrt{N}}\sum_{j=1}^{N}\left(\hat{a}_0 e^{i\delta \tau}+\hat{a}_0^{\dagger}e^{-i\delta \tau}\right)\hat{\sigma}^z_j\, ,
\end{align}
where $\Omega_0$ is proportional to the ODF and $\hat{a}_0$($\hat{a}_0^\dagger$) is the annihilation(creation) operator for the COM mode phonons.
In general, the spins will be coupled to the phonon mode, except at particular decoupling times $\tau_n = 2\pi n/\delta$ for an integer $n$ (Fig.~\ref{fig:expt}c and Supplementary Information).
Here we always choose $|\delta| = 2\pi n/\tau$, ensuring spins and phonons decouple.
This guarantees that the dynamics matches that of the Ising Hamiltonian in Eq.~\eqref{eq:Hspin} with uniform couplings $J(\delta)/\hbar=\Omega_0^2/(2\delta)$ and leads to different values of the coupling constant $J$ at different interaction times $\tau$.
The detuning-dependent coupling enables us to implement a many-body echo of the spin dynamics by inverting the sign of $\delta$.

For measuring magnetization and fidelity, we collect the global ion fluorescence scattered from the Doppler cooling laser on the cycling transition for ions in $\ket{\uparrow}$, after applying a $\pi/2$ rotation of the spins.
We count the total number of photons collected on a photomultiplier tube (PMT) in a detection period, typically $t_c=5\,$ms. 
From the independently calibrated photons collected per ion, we can infer the state populations, $N_\uparrow$ and $N_\downarrow$.
After averaging over many experimental trials, between 500 and 800, we calculate the expectation values $\langle{\hat S_z}\rangle = \langle{\hat{N}_\uparrow}\rangle - N/2$.
To measure the fidelity, we distinguish the single state with all ions in $\ket{\downarrow}$, which does not scatter from the cooling laser, from all other states. Any ion fluorescence indicates the system is no longer in the initial state. The fidelity is the fraction of experimental trials that result in measuring the state $\ket{\downarrow...\downarrow}$ (Supplementary Information).

\begin{figure}
    \centering
    \includegraphics[width=0.5\textwidth]{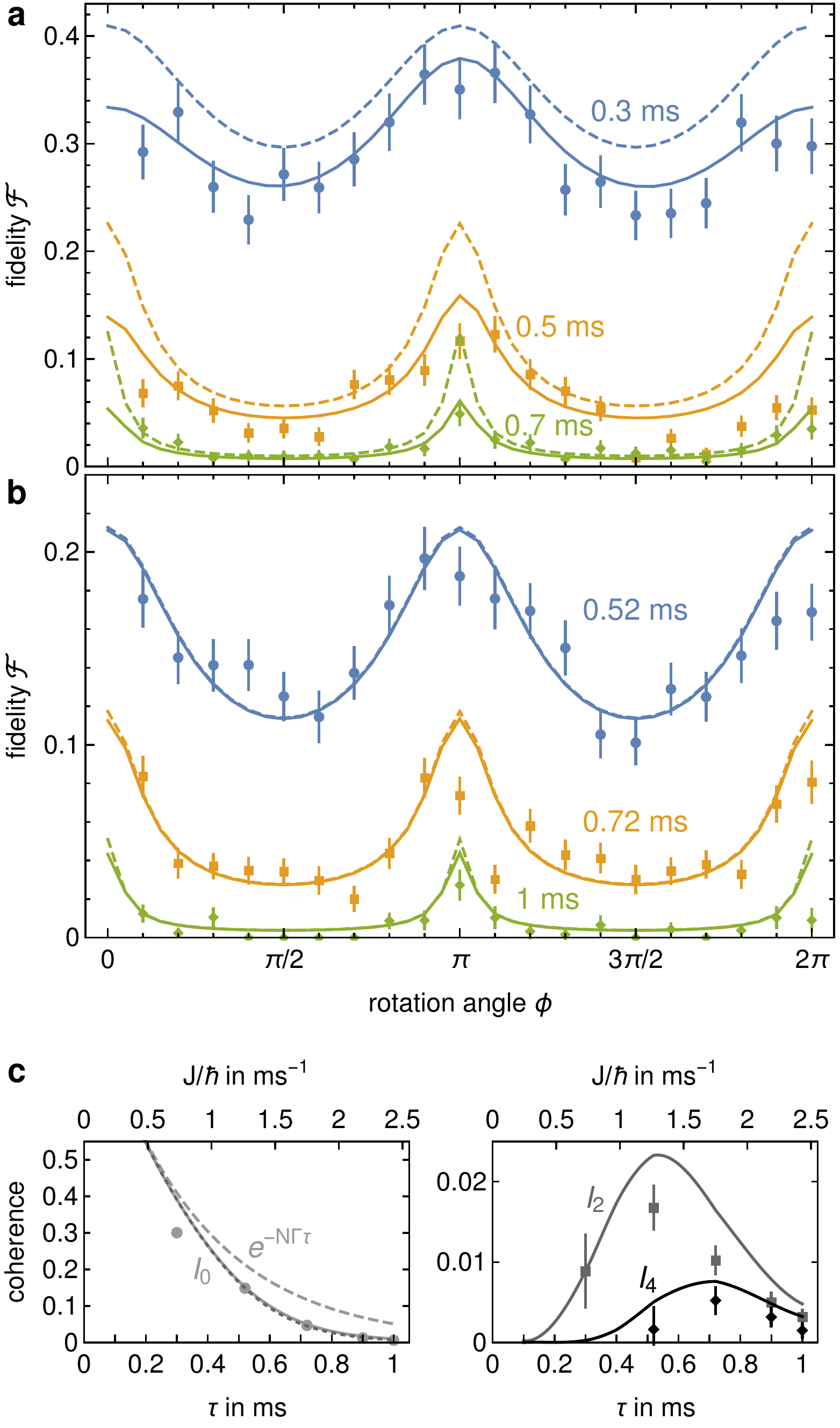}
    \caption{\textbf{Measured fidelity and coherence spectrum of $N=48$ ions.}
    \textbf{a},\textbf{b}, Dependence of the fidelity $\mathcal{F}_\phi(\tau)$ on the rotation angle $\phi$. The experimental sequence in \textbf{b} includes an additional $\pi$ pulse in the middle of each evolution period $\tau$. The dashed lines are simulations including off-resonant light scattering as the only source of decoherence, with $\Gamma=62\,$s$^{-1}$. The solid lines include effects of COM mode and magnetic field fluctuations, with COM mode frequency fluctuations $\Delta\ind{COM}/\omega_z=8.0\times10^{-5}$\,RMS, and magnetic field noise $\Delta_B/B = 0.32 \times 10^{-9}\,$ RMS (Methods). Note that for each interaction time $\tau$ the detuning is chosen so that $\delta = 2\pi/\tau$ (\textbf{a}) or $\delta = 4\pi/\tau$ (\textbf{b}). In each case, the spin-spin coupling also varies as $J/\hbar = \Omega_0^2/(2\delta)$\, where $\Omega_0 = 7850$~s$^{-1}$.
    \textbf{c}, Fourier amplitudes of fidelity (\textbf{b}) as a function of time. Solid lines are simulations including all known decoherence processes. $I_2$ and $I_4$ clearly show the buildup of higher order MQCs. Odd coherences and coherences $m\geq 6$ are zero within the statistical error. For $I_0$, decoherence induced decay (dashed) and approximate analytic curve (dotted, see text) are shown. The data points at $\tau=0.3$ and $0.9$ (not shown in \textbf{b}) have been added. The longest measured evolution time of $\tau=1\,$ms corresponds to $6.5\%$ of $t\ind{cat}$ (cf.\ Fig.~\ref{fig:seq}d).
    All error bars denote the statistical error of $1$ standard deviation (SD) of the mean.
    }
    \label{fig:F}
\end{figure}

Figure~\ref{fig:F} shows the measured fidelity $\mathcal{F}$ as a function of the angle $\phi$ for different evolution times $\tau$ in an array of 48 ions.
The measurements at $\phi=0$ and $2\pi$ give the state purity, while the $\pi$-periodic oscillations encode information on the buildup of MQCs.
The pulse sequence in Fig.~\ref{fig:F}a follows Fig.~\ref{fig:seq}, whereas in Fig.~\ref{fig:F}b, an additional $\pi$-rotation has been inserted in the middle of each evolution period $\tau$ to suppress some forms of decoherence.
We extract the coherences $I_m$, shown in Fig.~\ref{fig:F}c, as the Fourier components of the fidelity in Fig.~\ref{fig:F}b.
We see a clear buildup of the two-body ($I_2$), and then four-body ($I_4$) coherences with increasing interaction time.
Odd components are zero within statistical error, consistent with the fact that the coherences are generated by the Ising interaction, which can be viewed as only flipping pairs of spins.

All the measurements are in good agreement with theory calculations (solid lines) that account for independently calibrated sources of decoherence.
Off-resonant light scattering is the dominant decoherence mechanism in the system. Because the fidelity measures a projection onto a single many-body state, it decays with a rate approximately $N\Gamma$, where $\Gamma$ is the single particle decoherence rate. 
This causes a fast decay of $I_0$ as $\exp(-N\Gamma\tau)$. However, Fig.~\ref{fig:F}c shows that $I_0$ decays as $\exp(-N\Gamma\tau)I_0^\text{(pure)}$ where the algebraic decay $I_0^\text{(pure)}\approx1/(1+J^2\tau^2)$ (see Supplementary Information, Sec.~3) signals the buildup of higher-order coherences seen also in the fully coherent case.
Other sources of decoherence include slow drifts in the magnetic field \cite{Britton2016} and COM mode frequency fluctuations, which the MQC can distinguish.
Figure~\ref{fig:F}a reveals the degree to which the COM axial mode phonons cannot be decoupled from the spins due to uncertainty in the COM mode frequency $\omega_z$. The impact of residual spin-phonon coupling, arising from fluctuations in $\omega_z$, is more pronounced at $\phi=\pi$ than $\phi=0$. In contrast, slow magnetic field noise causes a reduction of the fidelity around $\phi= 0(2\pi)$, but has no effect at $\phi =\pi$, allowing us to benchmark these two imperfections independently.
For the data in Fig.~\ref{fig:F}b, where the sequence includes an additional $\pi$ rotation to suppress errors from slow drifts in the magnetic field and COM mode frequency, the full theory collapses to a solution that only includes off-resonant light scattering as the sole decoherence mechanism (dashed line).

\begin{figure}
    \centering
    \includegraphics[width=0.5\textwidth]{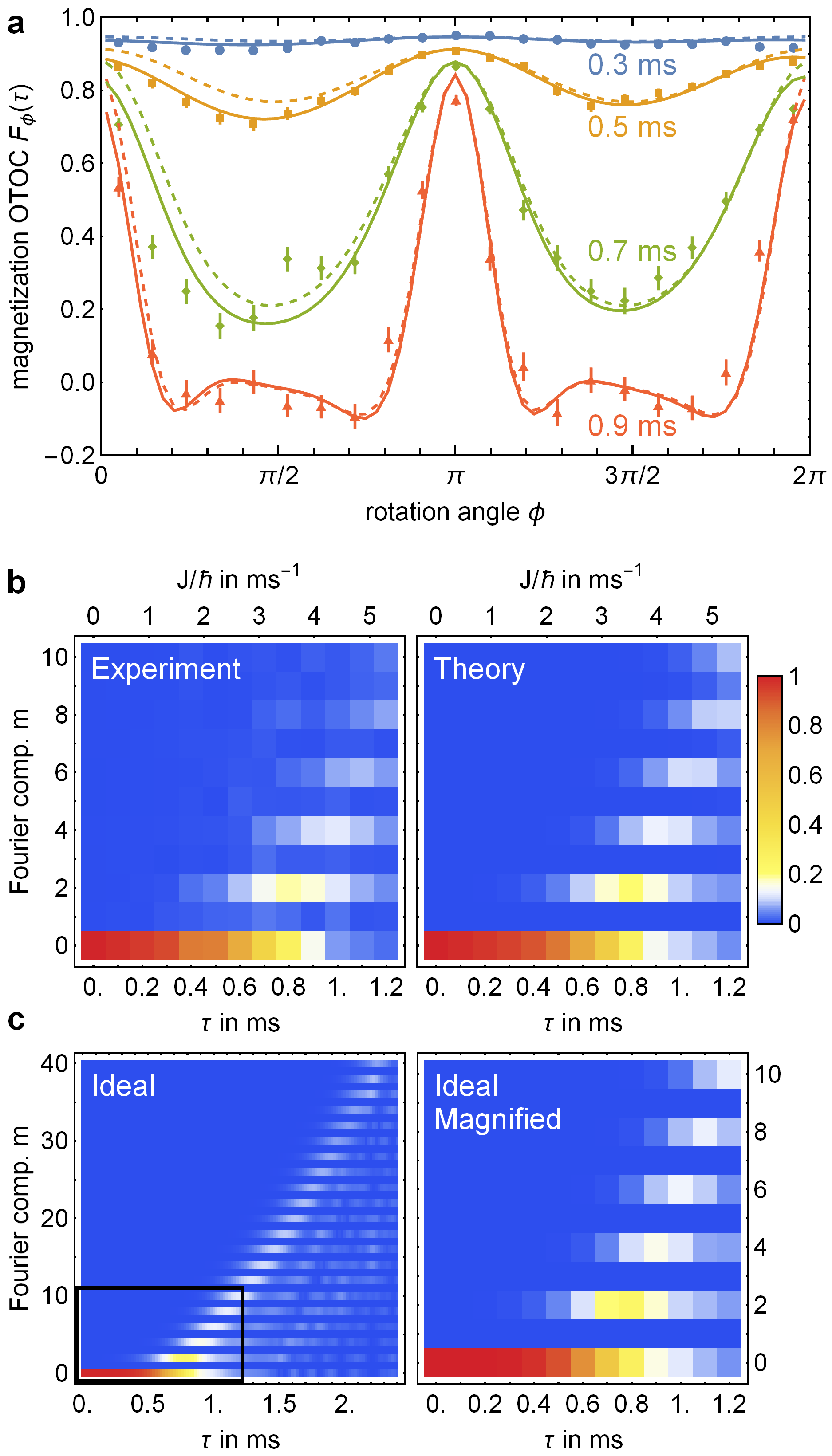}
    \caption{\textbf{Probing scrambling through magnetization dynamics.}
    \textbf{a}, Dependence of the normalized component $F_\phi(\tau)=(2/N)\langle \hat S_x\rangle$ of the total spin on the rotation angle $\phi$, measured in an array of $N=111(2)$ ions. Lines are the solutions of the full master equation with (solid) and without (dashed) magnetic field noise, where $\Delta_B/B = 0.32 \times 10^{-9}\,$ RMS. The effect of COM mode fluctuations is negligible here.
    Error bars denote the statistical error of $1$ SD of the mean.
    \textbf{b}, Fourier amplitudes  $A_m$ as a function of time. In the theory plot, the case without magnetic field noise (dashed lines in \textbf{a}) was used. The interaction parameter varies as $J/\hbar = \Omega_0^2/(2\delta)$\, where $\Omega_0 = 7450$ s$^{-1}$ and $\Gamma=91\,$s$^{-1}$. The longest measured evolution time of $\tau=1.2\,$ms corresponds to $7.3\%$ of $t\ind{cat}$.
    \textbf{c}, Ideal case for $N=111$, neglecting all decoherence effects. This corresponds to the lower panel of Fig.~\ref{fig:seq}d. The box in the left panel shows the experimentally accessed region which is magnified in the right panel.
    }
    \label{fig:Sz}
\end{figure}

Single-body observables, like the collective magnetization, are much less sensitive to decoherence, and provide an alternative way to experimentally measure the sequential buildup of higher order correlations induced by  spin-spin interactions. 
In Fig.~\ref{fig:Sz}, we show the results of the magnetization OTOC measurement sequence, which shows a buildup of Fourier amplitudes, $A_m$, up to $m=8$, observable even for $N=111$.
These measurements also allow us to benchmark the quality of our quantum simulator by comparing to theory predictions with no adjusted parameters.
Here, the dashed lines are obtained by solving the pure spin model including only spontaneous emission decoherence (see Supplementary Information) showing agreement in both the $\phi$-dependent signal (Fig.~\ref{fig:Sz}a) and its Fourier transform (Fig.~\ref{fig:Sz}b).
Accounting for static magnetic field noise largely explains remaining discrepancy at small angles (solid lines in Fig.~\ref{fig:Sz}a).
Comparison of the data to theory predictions with no decoherence (Fig.~\ref{fig:Sz}c) confirms that the decay of the Fourier amplitudes at long times is not a decoherence effect but a consequence of many-body interactions which induce a decrease of low-$m$ components with a corresponding buildup of high-$m$ components. Since the observed dynamics is dominated by the coherent evolution under the Ising interaction, these results suggest that the observed features can only be explained by the formation of quantum correlations.

In summary, we have shown that many-body Loschmidt echo sequences are powerful tools to measure OTOCs and quantify the degree of coherence in quantum simulators, with an explicit demonstration for ions in a Penning trap. In particular, we studied OTOCs involving variable angle spin rotations. The Fourier components with respect to the rotation angle ($I_m$ and $A_m$) show a buildup of many-body coherence and correlations, indicating scrambling of quantum information. Our experimental results are well described by a theory model which accounts for all known sources of decoherence (photon scattering, magnetic field noise, and spin-phonon coupling), allowing us to benchmark the performance of our trapped ion quantum simulator. 

The characteristic features of $A_m$s reported in this work demonstrate high level of control over the coherent many-body dynamics achieved by our trapped-ion quantum simulator and are fully consistent with the buildup of quantum correlations. Although currently the latter can be only indirectly inferred from the measurements, it is supported by previous benchmarking of the system using standard entanglement witnesses such as spin squeezing \cite{Bohnet2016}. We expect future work to derive formal connections between entanglement and scrambling, and to construct strict bounds that witness entanglement directly from $I_m$ and $A_m$ measurements.
While the current experimental system realizes a model amenable to classical simulations, we envision experiments going beyond this limit, e.g.\ by adding a spatially inhomogeneous magnetic field or preparing the system in non-symmetric or impure initial states, such as thermal states. These generalizations will allow us to explore the dynamics of OTOCs and characterize scrambling in unexplored regimes and under conditions where fast scrambling can occur.
Furthermore, the ability to time-reverse the dynamics will allow enhanced phase estimation without single particle detection resolution \cite{Davis2016, Macri2016, Linnemann2016}, investigations of quantum phase transitions \cite{Cucchietti2007}, criticality \cite{Quan2006}, thermalization in nearly closed quantum systems \cite{Eisert2015, Nandkishore2015} and the exploration of the  quantum-classical boundary \cite{Jacquod2009}, e.g., observation of the violation of Leggett-Garg inequalities \cite{Frowis2016}. 

After the completion of this work, we became aware of measurements of OTOCs using $4$ spins in an NMR system \cite{Li2016}.

\vspace*{0.5cm}
\begin{center}
  \textbf{METHODS}
\end{center}
\vspace*{0.5cm}

\noindent\textbf{Trap parameters.}
The experimental system is a two dimensional (2D) Wigner crystal of $^9$Be$^+$ ions formed in a Penning trap, described in Ref.~\cite{Bohnet2016}.
Details relevant to this work are documented here.
Axial confinement in the trap is provided by electric potentials applied to a stack of cylindrical electrodes; radial confinement is achieved using the Lorentz force produced by the controlled ion crystal rotation through the axial magnetic field, $B = 4.46\,$T, of the trap.
The axial trap frequency is $\omega_z = 1.570$ MHz, with a rotation frequency of $\omega_r = 180$ kHz.
The ions are laser cooled to the Doppler limit ($\approx 0.5\,$mK, with mean thermal occupation $\bar{n}\approx6$) using a pair of beams tuned close to the optical cycling transition $\ket{^2S_{1/2}J=+1/2, m_J=1/2} \rightarrow \ket{^2P_{3/2} J=+3/2,m_J=3/2}$.
The center of mass (COM) motional mode frequency is the trap axial frequency $\omega_z$, with the shorter wavelength modes well-resolved at lower frequencies \cite{Sawyer2012}.

\noindent\textbf{Qubit parameters.}
The two level spin system is formed by the valence electron spin in the magnetic field of the Penning trap, with $\SU = \ket{m_J = 1/2}$ ($\SD = \ket{m_J = -1/2}$) parallel (anti-parallel) to the field.
The spin states are split by 124\,GHz.
Global spin state rotations are performed using resonant microwaves, characterized by a $\pi$ rotation time of $70$\,$\mu$s.
The coherence time is primarily limited by shot-to-shot magnetic field fluctuations, causing fluctuations of the qubit state splitting with a standard deviation of 40 Hz.

\noindent{\bf Effective Hamiltonian parameters.}
The Ising Hamiltonian evolution is implemented using the spin-motion coupling described in Eq.~\eqref{eq:HSP}.
The spin-dependent ODF, with magnitude 
$|F_0|=|\hbar\Omega_0|/z_0$, where $z_0 \equiv \sqrt{\hbar/(2 m \omega_z)}$ is the ground state wave function size for a single trapped ion,
is provided by a pair of far-detuned 313 nm laser beams that excite the axial drumhead modes of the ion crystal.
The Coulomb force mediates effective spin-spin interactions through the spin-dependent motion, leading to well-characterized Ising interactions \cite{Britton2012, Bohnet2016}, similar to those used in a number of trapped ion quantum simulators  \cite{Porras2004,Richerme2014,Jurcevic2014}.
For this work, $|F_0|$ is typically 45\,yN.
The detuning $\delta$ in this work ranges from $2\pi\times$1 kHz to $2\pi\times$5\,kHz, and the impact of coupling to other axial modes, separated by at least $2\pi\times$25\,kHz from $\omega_z$, is negligible. 
For $\delta = 2\pi\times 1 \rm{kHz}$, we typically achieve $J/\hbar$ between 4000 and 5000 s$^{-1}$,  calibrated using mean field spin precession \cite{Britton2012} and collective spin depolarization \cite{Bohnet2016}.

\noindent{\bf Decoherence.}
The ODF beams scatter off-resonant photons, which is the primary source of decoherence during the coherent evolution.
We independently determine the total single particle decoherence rate $\Gamma$ using measurements of the decay of the second order moment of the collective spin $\langle \vec{S}^2 \rangle$.
Typically, $\Gamma \sim 65$\,s$^{-1}$, in good agreement with the prediction from the laser intensity alone.
However, for the data in Fig.~\ref{fig:Sz}, we measured $\Gamma \sim 90$\,s$^{-1}$, which we attribute to effects of the larger Lamb-Dicke confinement parameter for this data (see Supplementary Information of Ref.~\cite{Bohnet2016}), which had tighter radial trapping parameters than the data in Fig.~\ref{fig:F} (and therefore lower frequency transverse modes).
The decoherence rate $\Gamma$ is dominated by elastic Rayleigh scattering, which is $3.9$ times the total inelastic Raman scattering rate \cite{Uys2010}.

Beyond spontaneous emission, we also observe effects of fluctuations in the trap axial frequency $\omega_z$. 
Any fluctuations in the COM frequency will adversely affect the final state fidelity since entanglement between the spin and motional degrees of freedom is present unless operating precisely at the decoupling times $\tau_n$.
In addition, errors in $\delta$ lead to different spin-spin couplings in the two halves of the echo sequence, leading to imperfect many-body echoes (Supplementary Information).
We independently measure the effective COM mode stability with an experimental sequence like that of Fig.~\ref{fig:seq} with $\phi = \pi$ and $\delta$ is nominally set to zero, measuring $\langle\hat S_x\rangle$ versus $\tau$.
From the decay of $\langle\hat S_x\rangle$, we find the effective RMS fluctuations in $\omega_{z}$ to be $2\pi\times$125(50) Hz.
The incorporation of the COM mode frequency fluctuations in our theoretical model is described in a later section. 

We measure shot-to-shot magnetic field fluctuations using the experimental sequence of Fig.~\ref{fig:seq} but with the ODF beams blocked and $\phi=0$. From the decay of $\langle \hat S_x \rangle$ vs $\tau$, we determined magnetic-field induced  fluctuations in the qubit frequency of 40 Hz RMS.  We note that these measured fluctuations are the same order of, but somewhat smaller than measured previously ($\sim 68\,$Hz \cite{Britton2016}).

\noindent{\bf Experimental readout.}
To measure the fidelity, we distinguish the single state with all ions in $\ket{\downarrow}$, which does not scatter from the cooling laser, from all other states.
Off-resonant repumping from the cooling laser limits the detection time, and so setting a photon count threshold based on the average photons collected per ion generally underestimates the fidelity $\mathcal{F}$.
We recover nearly all the fidelity using a reference photon count distribution where all ions are prepared in $\ket{\downarrow}$ (Supplementary Information).

\noindent{\bf Multiple quantum coherence protocol.}
To prove the relation between the Fourier components of the fidelity and the  multiple quantum intensities (or coherences) $I_m$ we introduce the canonical product basis $\ket{\boldsymbol{\alpha}}=\ket{\alpha_1 \ldots \alpha_N}$, where $\alpha_i\in \{+,-\}$ and $\ket{\pm}$ are the eigenstates of $\hat \sigma_x$. The states $\ket{\boldsymbol{\alpha}}$ are eigenstates of $\hat S_x=\sum_i \hat \sigma_i^x/2$ with eigenvalues $M_x$, which are (half) integer numbers between $-N/2$ and $N/2$. We can now write the state $\hat\rho(\tau)$ as a sum of different coherence sectors $\hat\rho(\tau)=\sum_m \hat\rho_m$. Here, $\hat\rho_m$ contains all density matrix elements that account for coherences between basis states $\ket{\boldsymbol{\alpha}}$ and $\ket{\boldsymbol{\alpha}^\prime}$ for which $M_x-M_x^\prime=m$, i.e., which differ in the number of spins in the $\ket{+}$-state by $m$.

With this, one defines the MQC spectrum known from NMR \cite{Baum1985}:
\begin{equation}
    I_m=\tr[\hat\rho_m\hat\rho_{-m}]=\sum_{M_x-M_{x^\prime}=m}|\hat\rho_{\boldsymbol{\alpha}\boldsymbol{\alpha}^\prime}|^2  \,.
\end{equation}
Noting that a rotation about $x$ only results in the $m^{\mathrm{th}}$ sector picking up a phase $-m\phi$, one finds
\begin{equation}
\label{eq:FtoMQCmeth}
\begin{aligned}
   \mathcal{F}_\phi(\tau)
  & = \tr[\hat\rho_0\hat\rho_f] = \tr[\hat\rho_0 \hat{\mathcal{U}}^\dagger R_x(\phi) \hat{\mathcal{U}}\hat\rho_0\hat{\mathcal{U}}^\dagger \hat{R}_x^\dagger(\phi) \hat{\mathcal{U}}] \\
 & = \tr[\hat\rho(\tau)\hat\rho_\phi(\tau)]= \tr\left[\sum_{m^\prime}\hat\rho_{m^\prime}\sum_m \hat\rho_{m} e^{-i m\phi}\right] \\
  & = \sum_m \tr[\hat\rho_{-m}\hat\rho_m] e^{-i m\phi}=\sum_m I_m e^{-i m\phi}  \, .
\end{aligned}
\end{equation}
where $\hat\rho_\phi(\tau)=\hat{R}_x(\phi)\hat\rho(\tau)\hat{R}_x^\dagger(\phi)$ and $\hat{\mathcal{U}}=\exp[-i\hat H\ind{zz}\tau]$, and we have used cyclic permutations under the trace. The equality $\mathcal{F}_\phi(\tau)=\tr[\hat\rho(\tau)\hat\rho_\phi(\tau)]$ shows that for $\phi=0$ the fidelity measures the purity of the state $\hat{\rho}(\tau)$. Equation \eqref{eq:FtoMQCmeth} still holds in the presence of specific types of decoherence present in our experiment (Supplementary Information).

\noindent{\bf Scrambling of quantum information from Ising models.} 
In this section we provide further insight on the scrambling of quantum information.
We show that the $m^{\mathrm{th}}$ Fourier component of $F_{\phi}(\tau)$ is non-zero only if
at least one expectation value of an $n$-point operator, with $n\geq m$, is non-zero. Details of this calculation can be found in the Supplementary Information.

In the main text we defined the magnetization OTOC as the $x$-magnetization per spin $F_{\phi}(\tau)=(2/N)\langle \hat S_x\rangle=1/N\sum_{i=1}^N \langle\hat\sigma_i^x\rangle$, which with permutation symmetry simplifies to $F_{\phi}(\tau)=\langle\hat\sigma_i^x\rangle$ (for any $i$).
We can express the global magnetization $\langle \hat{S}^x\rangle$ at the end of the time reversal scheme in terms of an expectation value in $\hat\rho(\tau)$ (analogous to equation \eqref{eq:FtoMQCmeth}): $\langle \hat S^x\rangle = \tr[\hat R_x(\phi)\hat{\mathcal{U}}\hat S^x \hat{\mathcal{U}}^\dagger \hat R_x^\dagger(\phi)\hat\rho(\tau)]$
where $\hat{\mathcal{U}}$ is the unitary evolution under the interaction Hamiltonian, $\hat R_x(\phi)$ generates the rotation of the spins about $x$. Thus $\langle \hat S^x\rangle$ has the form $\langle e^{-i\hat S^x\phi}\hat O(\tau)e^{i\hat S^x\phi}\rangle_{\tau}$ where $\langle \cdot\rangle_{\tau}$ denotes the expectation value in state $\hat\rho(\tau)$. The general Hermitian operator $\hat O(\tau)$ can be written as a sum of products of single-spin operators $\hat O(\tau) = \sum_k a_k \prod_{j\in D_k} \hat{\sigma}_j^{b_j^k}$, 
where $D_k\subset\{1,\ldots,N\}$ is a set of particle indices and $b_j^k\in \{x,y,z\}$.
Applying the $x$-rotation to the operator $\hat O(\tau)$ is accomplished by replacing all Pauli spin operators according to: $\hat\sigma^x\rightarrow \hat\sigma^x$, $\hat\sigma^y\rightarrow \hat\sigma^y \cos(\phi) + \hat\sigma^z \sin(\phi)$, and $\hat\sigma^y\rightarrow \hat\sigma^z \cos(\phi) - \hat\sigma^y \sin(\phi)$. The resulting operator can be restated as
\begin{equation}
\begin{aligned}
    \langle e^{-i\hat S_x\phi}\hat O(\tau)e^{i\hat S_x\phi}\rangle_{\tau} &= \left\langle\sum_k \tilde{a}_k (\cos\phi)^{p_k}(\sin\phi)^{q_k}\prod_{j\in \tilde{D}_k} \hat\sigma_j^{\tilde{b}_j^k} \right\rangle_{\tau} \\
    & = \sum_k \tilde{a}_k (\cos\phi)^{p_k}(\sin\phi)^{q_k} \langle \hat{\mathcal{C}}_{\tilde{D}_k} \rangle_{\tau} \,.
\end{aligned}
\end{equation}
Here the tilde indicates that the coefficients and indices are different from the ones of $\hat O(\tau)$. $p_k$ and $q_k$ satisfy $0\leq p_k+q_k \leq N$. The crucial step is now to notice that terms with $p_k+q_k=m$ are associated with at least $m$-spin correlation functions, i.e. expectation values $\langle\hat{\mathcal{C}}_{\tilde{D}_k} \rangle$ of products of $|D_k|\geq m$ spin Pauli spin operators. Expanding $\langle e^{-i\hat S_x\phi}\hat O(\tau)e^{i\hat S_x\phi}\rangle$ into a Fourier series we find that terms with $p_k+q_k=m$ only contribute to Fourier components $A_n$ where $n\leq m$. Thus, if all correlation functions of more than $m$ spins are zero, then also all Fourier components above $m$ necessarily vanish. Conversely, if we observe a Fourier component $|A_m|>0$, we know that $n$-body correlations with $n\geq m$ must exist.  

In the Supplementary Information we illustrate this argument by considering the concrete case of an Ising model with arbitrary couplings $J_{ij}$. We show that for a $k$-local Ising model, in which any spin interacts with at most $k$ others, only Fourier components $A_m$ with $m\leq k+1$ can appear. For example in a one-dimensional Ising chain with nearest neighbor interactions no higher components than $A_3$ can buildup. In addition, the successive buildup of higher $A_m$ as a function of time $\tau$ and the vanishing of odd components can be understood in this way.

The observation of higher order Fourier components of $F_{\phi}(\tau)$ also allows to draw conclusions about the unitary evolution that creates the interacting dynamics. If this unitary is fully separable, i.e. a product of unitaries acting on single spins, such as a simple rotation, then $F_{\phi}(\tau)$ can at most develop Fourier components $|m|\leq 1$, since in this case $\langle S_x\rangle$ can be written as a sum of independent single-particle expectation values. This argument can be generalized to unitaries being products of terms acting on disjoint subsets of spins of at most size $\leq m$. If this is the case, then $A_{n>m}=0$. Thus, the observation of nonzero $A_m$ implies that the interaction Hamiltonian couples clusters of at least $m$ spins. We emphasize that this result is general, and does not rely on the assumption of an Ising interaction.

In the main text we noted that $\langle|[\hat{W}(\tau),\hat{V}]|^2\rangle=2(1-\text{Re}[F(\tau)])$.
This holds if both $\hat{W}(\tau)$ and $\hat{V}$ are unitary (as for $\hat{W}(\tau)=\hat{W}_\phi(\tau)=\exp(i\hat{H}\tau)\hat{R}_x(\phi) \exp(-i\hat{H}\tau)$ and $\hat{V}=\hat \sigma_i^x$). 
If $\hat{V}=\hat \rho_0=\ket{\psi_0}\bra{\psi_0}$ is the projector in the (pure) initial state, we obtain
\begin{equation}
    \langle|[\hat{W}(\tau),\hat{V}]|^2\rangle = 1-\langle \hat{W}^\dagger(\tau)  \hat{V} \hat{W}(\tau) \hat{V}\rangle = 1-\mathcal{F_\phi(\tau)}
\end{equation}
where we used that $\hat{V}=\hat{V}^\dagger$, $\hat{V}^2=\hat{V}$ and $\hat{V}\ket{\psi_0}=\ket{\psi_0}$.

\noindent{\bf Data availability.} The data that support the plots within this paper and other findings of this study are available from the corresponding author upon reasonable request.

\onecolumngrid

\vspace*{1cm}

\renewcommand{\theequation}{S\arabic{equation}}
\setcounter{equation}{0}

\begin{center}
 \textbf{Supplementary Information}
\end{center}
\vspace*{.5cm}

Here we provide details on the experimental readout techniques and elaborate on the connection between the buildup of many-body correlations and the Fourier components of the magnetization. Also, we give a detailed derivation of the approximate analytical expression for $I_0$ given in the main text. Finally, we specify the numerical methods used for including various decoherence effects. 

\section{Maximum likelihood estimation of the all-dark fraction}

\begin{figure}
    \centering
    \includegraphics[width=0.5\textwidth]{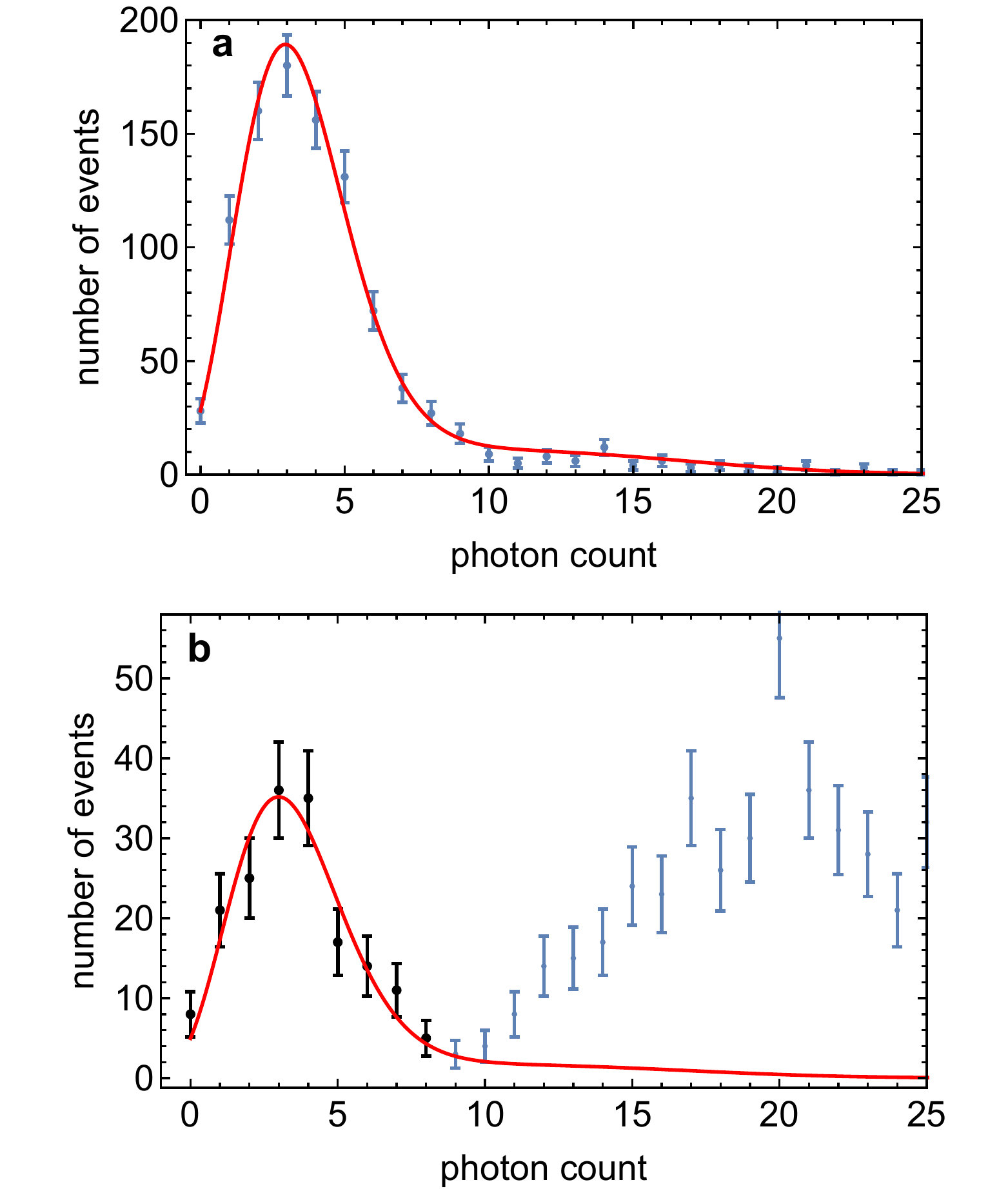}
    \caption{\textbf{Extraction of fidelity.} \textbf{a}, Reference histogram taken with all ions optically pumped into the "dark" state $\ket{\downarrow}$. \textbf{b}, Example of a photon count histogram obtained with the full MQC sequence.}
    \label{fig:ref_hist} 
\end{figure}

We improve our fidelity measurement by using a reference photon count distribution where all ions are prepared in $\ket{\downarrow}$ with a fidelity $>99.9\%$ \cite{Biercuk2009} to calibrate the background count rate $\Gamma_{d}$ and probability for a spin flip in the detection time $p\ind{flip}$ (Fig.~\ref{fig:ref_hist}a).
Then the expected count distribution is 
\begin{equation}
\begin{aligned}
    C(k) =& (1-p\ind{flip})P(\Gamma_d t_c,k) + \frac{p\ind{flip}}{t_c}  \int_0^{t_c} dt \sum_m P(\Gamma_d t,m)P[(\Gamma_d+\Gamma_b)(t_c-t),k-m] \\
     =& (1-p\ind{flip})P(\Gamma_d t_c,k) + p\ind{flip}  \frac{\Gamma(k+1,\Gamma_d t_c)+\Gamma[k+1,(\Gamma_d+\Gamma_b)t_c]}{\Gamma_b t_c k!}\, ,
\end{aligned}
\end{equation}
where $\tau$ is the detection time, $\Gamma_b$ is the independently calibrated scattering rate for an ion in the bright state, $P(\mu,k)$ is the Poisson distribution and $\Gamma(k,z)$ is the incomplete gamma function.
The probability for two spin flips is typically $\lesssim 1\%$, so we neglect these in our model.

Fitting the photon count histogram from each experiment can then extract the fidelity $\mathcal{F}$.
The amplitude of the peak determines the fidelity $\mathcal{F}$ for all spins being in $\ket{\downarrow}$. 
Fig.~\ref{fig:ref_hist}(b) shows a typical photon count histogram with the corresponding fit. 
For the fit, only the bins below a certain threshold are used to ensure that no events with one spin in the state $\ket{\uparrow}$ are counted. 
The threshold is chosen such that the expected contribution of events with one ion in the "bright" state $\ket{\uparrow}$ to the fitted bins is negligible. 

\section{Buildup of $m$-body correlations}

We showed in the main text that following the time-reversal protocol a measurement of the single particle operator $\langle \hat \sigma_i^x \rangle$ is an out-of-time-order correlator $F_\phi(\tau)$ and quantifies the scrambling of quantum information. In this section we show that the observation of higher order Fourier components $A_m$ of  $F_\phi(\tau)$, indicates the buildup of higher order correlations between the spins. We illustrate how this leads to a better understanding of the features observed in Fig.~4.

\subsection{General argument.}

By measuring the global magnetization $\langle \hat{S}^x\rangle$ at the end of the time reversal scheme we obtain (similar to equation (8))
\begin{equation}
    \langle \hat S^x\rangle = \tr[\hat S^x \hat\rho_f] = \tr[\hat S^x \hat{\mathcal{U}}^\dagger \hat R_x^\dagger(\phi)\hat\rho(\tau)\hat R_x(\phi)\hat{\mathcal{U}}] = \tr[\hat R_x(\phi)\hat{\mathcal{U}}\hat S^x \hat{\mathcal{U}}^\dagger \hat R_x^\dagger(\phi)\hat\rho(\tau)]
\end{equation}
where $\hat{\mathcal{U}}$ is the unitary evolution under the interaction Hamiltonian, $\hat R_x(\phi)$ generates the rotation of the spins about $x$ and $\hat\rho(\tau)$ is the time evolved (under $\hat{\mathcal{U}}$) state that we are interested in. Thus $\langle \hat S^x\rangle$ has the form $\langle e^{-i\hat S^x\phi}\hat O(\tau)e^{i\hat S^x\phi}\rangle_{\tau}$ where $\langle \cdot\rangle_{\tau}$ now denotes the expectation value in state $\hat\rho(\tau)$. The general Hermitian operator $\hat O(\tau)$ can be written as a sum of products of single-spin operators
\begin{equation}
    \hat O(\tau) = \sum_k a_k \prod_{j\in D_k} \hat{\sigma}_j^{b_j^k} 
\end{equation}
where $D_k\subset\{1,\ldots,N\}$ is a set of particle indices and $b_j^k\in \{x,y,z\}$. This representation exploits the fact that all possible products of $N$ operators with factors in $\{\hat 1,\hat\sigma^x,\hat\sigma^y,\hat\sigma^z\}$ form a complete set of operators acting on the Hilbert space of $N$ spins.
Applying the $x$-rotation to the operator $\hat O(\tau)$ is accomplished by replacing all Pauli spin operators according to: $\hat\sigma^x\rightarrow \hat\sigma^x$, $\hat\sigma^y\rightarrow \hat\sigma^y \cos(\phi) + \hat\sigma^z \sin(\phi)$, and $\hat\sigma^y\rightarrow \hat\sigma^z \cos(\phi) - \hat\sigma^y \sin(\phi)$. After doing this the products can be multiplied out and the sum reordered, yielding
\begin{equation}
    \langle e^{-i\hat S_x\phi}\hat O(\tau)e^{i\hat S_x\phi}\rangle_{\tau} = \left\langle\sum_k \tilde{a}_k (\cos\phi)^{p_k}(\sin\phi)^{q_k}\prod_{j\in \tilde{D}_k} \hat\sigma_j^{\tilde{b}_j^k} \right\rangle_{\tau} = \sum_k \tilde{a}_k (\cos\phi)^{p_k}(\sin\phi)^{q_k} \langle \hat{\mathcal{C}}_{\tilde{D}_k} \rangle_{\tau} \,.
\end{equation}
Here the tilde indicates that the coefficients and indices are different from the previous ones. $p_k$ and $q_k$ are non-negative integers satisfying $0\leq p_k+q_k \leq N$. The crucial step is now, to notice that terms with $p_k+q_k=m$ are associated with at least $m$-spin correlation functions, i.e. expectation values $\langle\mathcal{C}_{\tilde{D}_k} \rangle$ of products of $|D_k|\geq m$ spin Pauli spin operators. Expanding $\langle e^{-iS_x\phi}O(\tau)e^{iS_x\phi}\rangle$ into a Fourier series we find that terms with $p_k+q_k=m$ only contribute to Fourier components $A_n$ where $n\leq m$. Thus, if all correlation functions of more than $m$ spins are zero, then also all Fourier components above $m$ necessarily vanish. Conversely, if we observe a Fourier component $|A_m|>0$, we know that $n$-body correlations with $n\geq m$ must exist.  

\subsection{Ising models with general coupling coefficients.}

In the above argument we made no assumptions about the interaction Hamiltonian or the observable that is measured. We now want to consider the concrete case of evolution under an Ising Hamiltonian and measurement of the magnetization $(N/2)F_\phi(\tau)=\langle S_x\rangle=1/2\sum_{i=1}^N \langle\sigma_i^x\rangle$.
We find that for Hamiltonians with pair-wise Ising interactions (i) the $m^{\mathrm{th}}$ Fourier component of $F_\phi(\tau)$ is non-zero only if at least one spin, labeled $i$, in the system is coupled to $m-1$ other spins in the system and that (ii) higher order Fourier components build up at higher order in the short-time expansion.

The system evolves under the Hamiltonian $\hat H_{\mathrm{zz}}=\sum_{j\neq k} J_{jk} \hat \sigma_j^z \hat\sigma_k^z$, which can be split into two parts: 
\begin{align}
\hat H_{\rm zz}&=\frac12 \hat B_{\rm eff}^i \hat\sigma_i^z + \hat{\tilde H}^i\, , \\
\hat{\tilde H}^i&= \sum_{j\neq k\neq i} J_{jk} \hat \sigma_j^z \hat \sigma_k^z\, ,\\
\hat B_{\rm eff}&= 4 \sum_{j\neq i} J_{ij} \hat \sigma_j^z\, .
\end{align}

\begin{figure}
    \centering
    \includegraphics[trim={0cm 3cm 17cm 0cm},clip, width=0.5\textwidth]{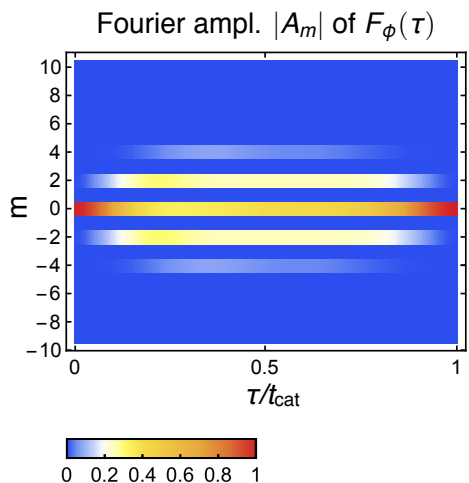}
    \caption{\textbf{Dynamics of Fourier components of the magnetization OTOC $F_\phi(\tau)$ in a 1D chain with $N=10$. a,} Under the nearest-neighbor Ising model each particle is only linked to two other particles and $m=\pm 2$ are the highest non-zero Fourier components observed. \textbf{b, } In the presence of next-nearest-neighbor Ising interactions there are four direct links for each particle and $m=\pm 4$ are the highest non-zero Fourier components observed.}
    \label{fig:correlations}
\end{figure}

Without loss of generality we calculate the expectation value $\langle \sigma_i^x\rangle$. All statements we make immediately generalize to the magnetization OTOC $F_\phi(\tau)=1/N\sum_{i=1}^N \langle\sigma_i^x\rangle$. Again evolving under the MQC sequence (assuming a pure state $\rho(\tau)=\ket{\psi}\bra{\psi}$ for convenience): 
\begin{equation}
\label{eq:sigx1}
    \langle \sigma_i^x\rangle
     = \bra{\psi} e^{-i \hat S^x \phi} e^{-i \hat H_{\rm zz}\tau}\hat \sigma_i^x
     e^{i \hat H_{\rm zz} \tau}e^{i \hat S^x \phi} \ket{\psi}
     =\bra{\psi} e^{-i \hat S^x \phi} e^{-i \hat B_{\rm eff}^i \tau/2}\hat \sigma_i^x 
    e^{i \hat B_{\rm eff}^i \tau/2}e^{i \hat S^x \phi}\ket{\psi} \, ,
\end{equation}
where we have used $[\sigma_i^x,\hat{\tilde H}^i] =0$.
Using the relationship
$$
e^{-i\hat B_{\rm eff}^i \tau/2} \hat \sigma_i^xe^{i\hat B_{\rm eff}^i \tau/2} = e^{-i\hat B_{\rm eff}^i \tau} \hat{\sigma}_i^++e^{i\hat B_{\rm eff}^i \tau} \hat{\sigma}_i^-\, ,
$$
we can rewrite equation~\eqref{eq:sigx1} as 
\begin{equation}
\label{eq:sigx2}
\langle \sigma_i^x\rangle =
 \left\langle 1/2[\sigma_i^x+i(\sigma_i^y\cos\phi+\sigma_i^z\sin\phi)]\hat{\tilde O}_i \right\rangle_\tau +\text{c.c.}
\end{equation}
where
\begin{equation}
\label{eq:O1}
\hat{\tilde O}_i=e^{-i \tau( \cos\phi \hat{\tilde S}_i^z-\sin\phi\hat{\tilde S}_i^y)}\, .
\end{equation}
Here, $\hat{\tilde S}_i^\alpha={\sum_{j=1}^m}^\prime 4 J_{ij} \hat\sigma_j^\alpha$ with $\alpha=y, z$ and the prime on the summation indicates that particle $i$ is excluded and $m$ is the number of particles connected to particle $i$ via interactions $J_{ij}$ (i.e. the sum over $j$ only includes non-zero terms $J_{ij}$). 
It is useful to expand equation~\eqref{eq:O1} in the form
\begin{equation}
\label{eq:O2S}
 \hat{\tilde O}_i=\prod_{j=1}^m \left(\cos(4J_{ij}\tau)\hat{I}_j-i \sin(4J_{ij}\tau) \left(\cos\phi\, \hat \sigma_j^z-\sin\phi\, \hat \sigma_j^y\right)\right)\, .
\end{equation}
From equations~\eqref{eq:sigx2} and \eqref{eq:O2S} we see that $\langle \hat{\sigma}_i^x \rangle$ has at most $(m+1)$ non-zero Fourier components, that is one more than the number of interactions links between particle $i$ and the rest of the system. This point is also illustrated in figure~\ref{fig:correlations}. Panels (a) and (b) correspond to one-dimensional chains with nearest-neighbor and next-nearest-neighbor couplings, respectively. Consequently in the former $\vert A_2\vert$ is the highest non-zero Fourier component, while in the latter $\vert A_4 \vert$ is also observed. This proves statement (i).

Next, we write an explicit expression for $F_\phi(\tau)$ in terms of $n$-point functions by expanding the exponential in equation~\eqref{eq:O1}: 
\begin{align}
\label{eq:O3S}
\nonumber \hat{\tilde O}_i &=e^{-i \tau (\cos\phi \hat{\tilde S}_i^z-\sin\phi\hat{\tilde S}_i^y)}\\
\nonumber &= \sum_{n=0}^\infty \frac{({i\tau})^n}{n!} \left[\cos\phi\,\hat{\tilde S}_i^z - \sin\phi\,\hat{\tilde S}_i^y\right]^n\\
&=\sum_{n=0}^\infty \frac{({i\tau})^n}{n!}\sum_{p=0}^n (\cos\phi)^p (\sin\phi)^{n-p}\, \hat{\mathcal{P}}_{p,n}\biggl\lbrace \hat{\tilde S}_i^z, \hat{\tilde S}_i^y\biggr\rbrace\, ,
\end{align}
where $\hat{\mathcal{P}}_{p,n}\lbrace \hat{A},\hat{B}  \rbrace$ is used to denote the equally weighted sum of all operators with $p$ $\hat{A}$s and $(n-p)$ $\hat{B}$s. For example, $\hat{\mathcal{P}}_{1,3}\{ \hat{\tilde S}_i^z, \hat{\tilde S}_i^y\}=\hat{\tilde S}_i^z (\hat{\tilde S}_i^y)^2+\hat{\tilde S}_i^y\hat{\tilde S}_i^z\hat{\tilde S}_i^y+(\hat{\tilde S}_i^y)^2 \hat{\tilde S}_i^z$.

Equations~\eqref{eq:sigx2} and \eqref{eq:O3S} together show that the amplitude of $m^{\mathrm{th}}$ Fourier component of $F_\phi(\tau)$, denoted by $A_{m}$, is determined by the magnitude of (at least) $m$-point functions of the form 
$\langle \hat\sigma^\alpha\hat{\mathcal{P}}_{p,n=m-1}\lbrace \hat{\tilde S}_i^z, \hat{\tilde S}_i^y\rbrace\rangle_\tau$, where the expectation is in the state $\rho$. 
Furthermore equation~\eqref{eq:O3S} shows that at short times $A_m$ grows as $\tau^{m-1}$ (or slower since the $n$-point functions depend on $\tau$ themselves and possibly vanish at $0$th order in the small $\tau$ expansion), which proves (ii). 

\subsection{All-to-all Ising dynamcis.}

If we consider the case of a permutation symmetric Ising model, which we study in the main text, the expression \eqref{eq:O3S} can be recast in terms of collective spin operators, giving
\begin{equation}
\begin{aligned}
     F_\phi(\tau)= \frac{2}{N}e^{-iJt/2} & \left\langle \hat S_x\sum_{\text{even }n} \frac{(iJt)^n}{n!} \sum_{p=0}^n \hat{\mathcal{P}}_{p,n}\lbrace \hat{S}^z,\hat{S}^y\rbrace\cos^p\phi \sin^{n-p}\phi \right. \\
     & \left. -i(\hat S_y\cos\phi - \hat S_z\sin\phi) \sum_{\text{odd }n} \frac{(iJt)^n}{n!} \sum_{p=0}^n \hat{\mathcal{P}}_{p,n}\lbrace \hat{S}^z,\hat{S}^y\rbrace\cos^p\phi \sin^{n-p}\phi \right\rangle_\tau \,.
\end{aligned}
\end{equation}
This again illustrates the sequential buildup of higher Fourier components $A_m$. We also note that the above expression is a sum of terms $\propto\cos^p\phi \sin^q\phi$, where $p+q$ is always even. Since for even $p+q$ these terms contribute only to even Fourier components $m\leq p+q$, this explains the observed absence of odd Fourier components of $F_\phi(\tau)$.

\section{Dynamics of $I_0$}

In this section we show that for pure states the $I_m$ can be related to the counting statistics of the spins $\lbrace P_n\rbrace$, where a specific $P_n$ denotes the probability of measuring a state with exactly $n$ particles in the $\vert + \rangle$ state. 

Consider the definition of  $I_m= \tr  \left[\hat \rho_m \hat \rho_{-m}\right]$. For pure collective states we can write  $I_m$ using $\vert \psi \rangle = \sum_{m} c_m  \vert m\rangle$ with $\hat S_x \vert m\rangle= (m-N/2) \vert m \rangle$ as, 
\begin{align}
I_m= \sum_n P_n P_{m+n}\, ,\label{eq:ImPn}
\end{align}
where $P_m=\vert c_m\vert^2$. 
This relationship allows us to derive an analytic expression for $I_0= \sum_n \vert P_n\vert^2$ for pure states. In the Dicke basis we fine that 
\begin{equation}
 P_n(\tau)=\biggl\vert \biggl(\sum_{m=0}^N\sum_{p=0}^n\binom{N}{m} \binom{m}{p}\binom{N-m}{n-p} (-1)^p e^{i2J/N(m-N/2)^2 \tau}\biggr)\bigg /\left(2^N \sqrt{\binom{N}{n}}\right)\biggr \vert^2
\end{equation}

\begin{figure}
    \centering
    \includegraphics[trim={0.5cm 15cm 19cm 0cm},clip, width=0.5\textwidth]{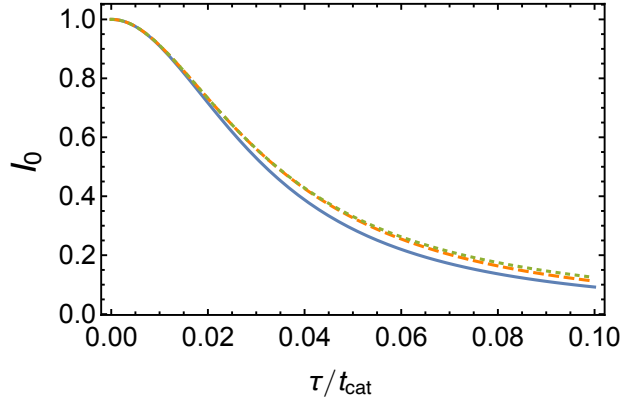}
    \caption{\textbf{Approximate dynamics of $I_0$.} The solid, dashed, and dotted lines use Eq.~\eqref{eq:ImPn} with an upper limit on the summation of 0, 1, and 2, respectively.}
    \label{fig:ImPm}
\end{figure}

In Fig.~\ref{fig:ImPm} we plot the above series for $I_0$, terminating the series at 1, 2, or 3 terms. It is evident that the dynamics of $I_0$ are to a large extent captured by $P_0^2$, since at short times, to leading order, $P_n(\tau)\propto \tau^{2n}$ for $n>0$. Moreover for experimentally relevant timescales one can obtain a simple analytic formula using the normal approximation to the binomial distribution, 
\begin{equation}
    I_0(\tau) \approx P_0^2(\tau)= \frac{1}{1+J^2\tau^2}.
\end{equation}

\section{Effect of decoherence on $F_\phi(\tau)$}

\begin{figure}
    \centering
    \includegraphics[width=0.5\textwidth]{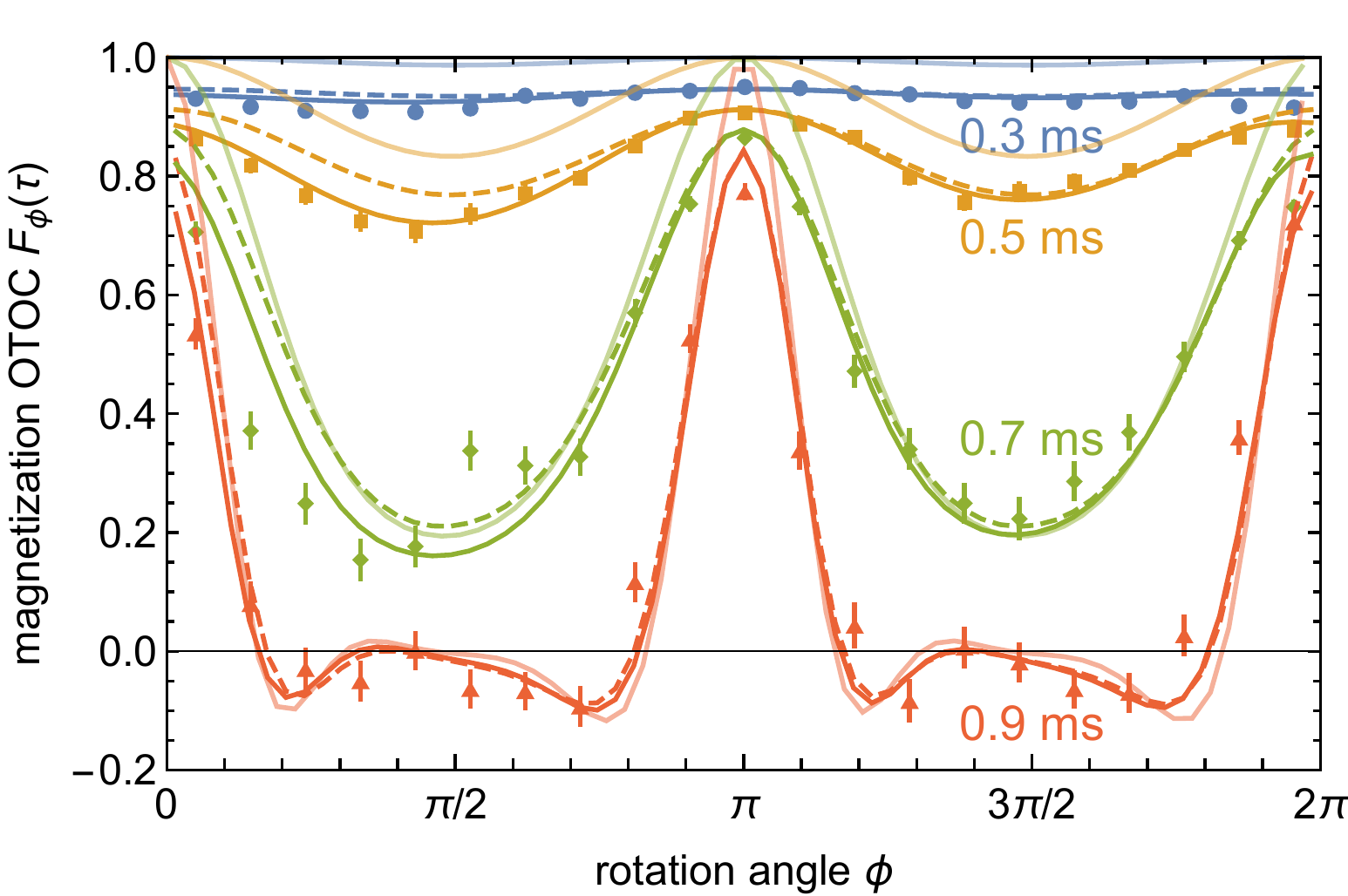}
    \caption{\textbf{Effect of decoherence on $F_\phi(\tau)$.} As in Fig.~4a, dashed and solid lines show the magnetization OTOC accounting for decoherence due to off-resonant light scattering and additionally static magnetic field noise, respectively. The faint solid lines show the ideal case neglecting all sources of decoherence.}
    \label{fig:decohSz}
\end{figure}

In order to illustrate more clearly the effects of decoherence due to off-resonant light scattering on the magnetization OTOC, we show Fig.~4 with the decoherence-free case added (faint lines in Fig.~\ref{fig:decohSz}). It can be seen that decoherence leads to a global decay of $F_\phi(\tau)$ for short times $\tau$, while at longer times the effects are less obvious.

\section{Solution of the full master equation}\label{sec:master}

The evolution including decoherence due to photon scattering is governed by a master equation of Lindblad form
\begin{equation}
\label{eq:ME}
 \dot{\hat\rho}=-i[\hat H,\hat\rho]+\sum_k \mathcal{L}(\hat{\Gamma}_k)\hat\rho\, ,
\end{equation}
where $\hat H=\hat H\ind{zz}+B\sum_i\hat \sigma_i^z$ (the second term is needed if static magnetic-field noise is included) and
\begin{equation}
 \mathcal{L}(\hat{\Gamma}_k)\hat\rho = \sum_i\hat{\Gamma}_{k,i}\hat\rho \hat{\Gamma}_{k,i}^\dagger  -\frac{1}{2}(\hat{\Gamma}_{k,i}^\dagger \hat{\Gamma}_{k,i}\hat\rho+ \hat\rho\hat{\Gamma}_{k,i}^\dagger \hat{\Gamma}_{k,i} )\, ,
\end{equation}
is the Lindblad superoperator ($i$ is the spin index). The relevant types of decoherence to consider are spontaneous emission up ($\hat{\Gamma}_{\text{du},i}=\sqrt{\Gamma_{\text{du}}}\ket{\uparrow_i}\bra{\downarrow_i}$) and down ($\hat{\Gamma}_{\text{ud},i}=\sqrt{\Gamma_{\text{ud}}}\ket{\downarrow_i}\bra{\uparrow_i}$) and dephasing [$\hat{\Gamma}_{\text{el},i}=\sqrt{\Gamma_{\text{el}}}/2(\ket{\uparrow_i}\bra{\uparrow_i}-\ket{\downarrow_i}\bra{\downarrow_i}$)] \cite{Uys2010}. 

All terms of the master equation are invariant under exchange of the particle indices. This symmetry can be exploited to drastically reduce the complexity of the problem. The density matrix of the initial state is fully symmetric under particle exchange and this symmetry is conserved by the time evolution. Within the $4^N$-dimensional space of $N$-spin density matrices the dynamics is therefore restricted to the subspace of fully symmetric density matrices, which has dimension $(N+1)(N+2)(N+3)/6$. Techniques to represent and solve the master equation on this symmetrized Liouville space have been discussed for example in Refs.~\cite{Sarkar1987, Hartmann2012, Xu2013}.
In the present case the Liouvillian is block diagonal with blocks of dimension $\sim N$. Thus the complexity of one time propagation step is $\sim N^4$. Rotations can be performed analytically and can be decomposed into block diagonal superoperators with the number of non-zero matrix elements of order $N^4$. This efficient implementation allows us to solve the master equation of more than $100$ of spins on a conventional computer. This method was used to produce the dashed and solid lines shown in Fig.~4 in the main text.

The Liouville picture also makes it possible to understand how the MQC scheme is affected by decoherence.
As long as $\Gamma\ind{ud}=\Gamma\ind{du}$, which is fulfilled to a good approximation, the superoperator of the Lindblad terms is diagonal and in this case equation~(4) of the main text still holds, meaning that the Fourier components of the fidelity still exactly represent the coherences $I_m$ of the state $\hat\rho(\tau)$. In particular, the equality $\mathcal{F}_\phi(\tau)=\tr[\hat\rho_0\hat\rho_f]=\tr[\hat\rho(\tau)\hat\rho_\phi(\tau)]$ still holds showing that $\mathcal{F}_0(\tau)$ measures the purity of $\hat \rho(\tau)$.

The bare values of $\Gamma\ind{ud}$, $\Gamma\ind{du}$, and $\Gamma\ind{el}$ are calculated based on the polarizations and a measurement of the intensity of the ODF beams. For higher rotation frequencies where the Lamb-Dicke confinement criterion is not well satisfied we measure an additional decay $\Gamma\ind{add}$ of the total spin vector as a function of the time the ODF beams are on.
The total decoherence rate (quoted in the main text) is obtained as $\Gamma=(\Gamma\ind{el}+\Gamma\ind{add}+\Gamma\ind{du}+\Gamma\ind{ud})/2$. The single spin coherences decay with this rate.

For the data presented in Fig.~3, the scattering rates calculated from laser intensity measurements are $\Gamma\ind{el}=91$\,s$^{-1}$, $\Gamma\ind{ud}=14$\,s$^{-1}$, $\Gamma\ind{du}=10$\,s$^{-1}$. Independent measurements are consistent with $\Gamma\ind{add}=0$ but have an uncertainty of about 10\%, i.e., $\Gamma=57(6)\,$s$^{-1}$. The fidelity measurement actually provides a more sensitive way to determine $\Gamma$, which motivates the choice of $\Gamma=62\,$s$^{-1}$ for the simulations shown in Fig.~3. By solving the full master equation as outlined above, we confirmed that at short times ($\tau \lesssim 1\,$ms) the decoherence can be accounted for by globally reducing the fidelity by a factor $\exp(-N\Gamma\tau)$.

For the simulations in Fig.~4 we used $\Gamma\ind{ud}=14$\,s$^{-1}$, $\Gamma\ind{du}=10$\,s$^{-1}$, $\Gamma\ind{el}=94$\,s$^{-1} +\Gamma\ind{add}$, $\Gamma\ind{add}=65$\,s$^{-1}$, where $\Gamma\ind{add}$ was determined in an independent measurement.

\section{Spin-motion coupling}

The trapped ion simulator utilizes spin-motion interactions to generate the effective spin-spin couplings. In the regime where only the COM mode participates, the system dynamics are generated by the interaction picture spin-phonon Hamiltonian given by
\begin{equation}
\label{eq:Hsupp}
\hat H_{\rm full} =-\frac{\Omega_0}{\sqrt{N}}\sum_{j=1}^N \cos (\mu_r t + \varphi) (\hat a_0 e^{-i\omega_z t}+ \hat a_0^\dagger e^{i\omega_z t})\hat \sigma_j^z\, ,
\end{equation}
where
% $\Omega_0=F_0\eta_0/k_R$, with $F_0$ the magnitude of the spin-dependent force, 
$\omega_z$ is the center of mass frequency, $\varphi$ is the ODF phase, and $\mu_r=\omega_z+\delta$ is the ODF beat note frequency, with $\delta$ denoting the detuning. The system dynamics is generated by the propagator $\hat{\mathcal{U}}=\hat{\mathcal{U}}_{\rm SP} \hat{\mathcal{U}}_{\rm SS}$ given by
\begin{align}
\hat{\mathcal{U}}_{\rm SP}(t,t_0)&=\exp\left[ \sum_j \left(\alpha (t,t_0) \hat{a}_0^\dagger - \bar\alpha (t,t_0) \hat{a}_0 \right) \hat \sigma_j^z\right]\, ,\\
\hat{\mathcal{U}}_{\rm SS}(t,t_0)&=\exp\left[-i J(t,t_0)  \sum_{i < j } \hat \sigma_i^z \hat \sigma_j^z\right]\, ,
\end{align}
where
\begin{align}
\label{eq:alphafull}
\alpha(t,t_0)&= i \frac{\Omega_0}{\sqrt N} \int_{t_0}^t d\tau e^{i \omega_z\tau}\cos(\mu_r \tau+\varphi),  \\
\label{eq:Jijfull}
J(t,t_0) &=\frac{2 \Omega_0^2}{N} \int_{t_0}^t d\tau \int_{t_0}^\tau d\tau^\prime  \cos(\mu_r \tau^\prime+\varphi)\cos(\mu_r \tau+\varphi)\\ 
&\hskip 40 pt 
\times \sin(\omega_z(\tau-\tau^\prime)), 
\end{align}
where $\bar\bullet$ denotes complex conjugation. In the regime where $\delta\ll \omega_z$ one may use the rotating wave approximation. Then the Hamiltonian $\hat H_{\rm full}$ reduces to $\hat H_I$ given by Eq.~(6) and the expressions for $\alpha$ and $J$ simplify to
\begin{align}
\alpha(t,t_0) &\approx-\frac{\Omega_0}{2\delta\sqrt N} e^{- i \varphi} \left(e^{- i \delta t}- e^{- i \delta t_0}\right)\\
J(t, t_0)&\approx \frac{\Omega_0^2}{2 N\delta}(t-t_0)\, ,
\end{align}
where in the last expression we provide the secular expression for $J(t, t_0)$. Within these approximate expressions it is easy to see that at times $t_n=t_0+2\pi n /\delta$, where $n$ is an integer, spin and motion decouple and the dynamics of the system resembles that of a spin system with uniform Ising interactions given by $J= \frac{\Omega_0^2}{2 N\delta}$.

The experimental sequence consists of an initial rotation aligning the spins such that they are pointing along the $x$-axis, 
$$\vert \psi (0)\rangle = \prod_{j=1}^N \left(\frac{1}{\sqrt 2} \vert \uparrow \rangle_j +\frac{1}{\sqrt 2}\vert \downarrow \rangle_j \right)\, , $$
The initial density matrix of the system including the thermal phonons is given by:
\begin{equation}
\hat\rho(0)=\vert \psi(0)\rangle \langle \psi(0) \vert\otimes \hat\rho_{\rm thermal} \, .
\end{equation}
The time evolution is generated by the propagator $\hat{\mathcal{U}}(2\tau, 0)=\hat{\mathcal{U}}(2\tau, \tau) \hat{R}_x(\phi) \hat{\mathcal{U}}(\tau, 0)$, where $\hat{R}_x(\phi)$ denotes a rotation around the $x$ direction by angle $\phi$. 

\begin{figure}
    \centering
    \includegraphics[trim={1cm 3cm 19cm 1cm},clip, width=0.5\textwidth]{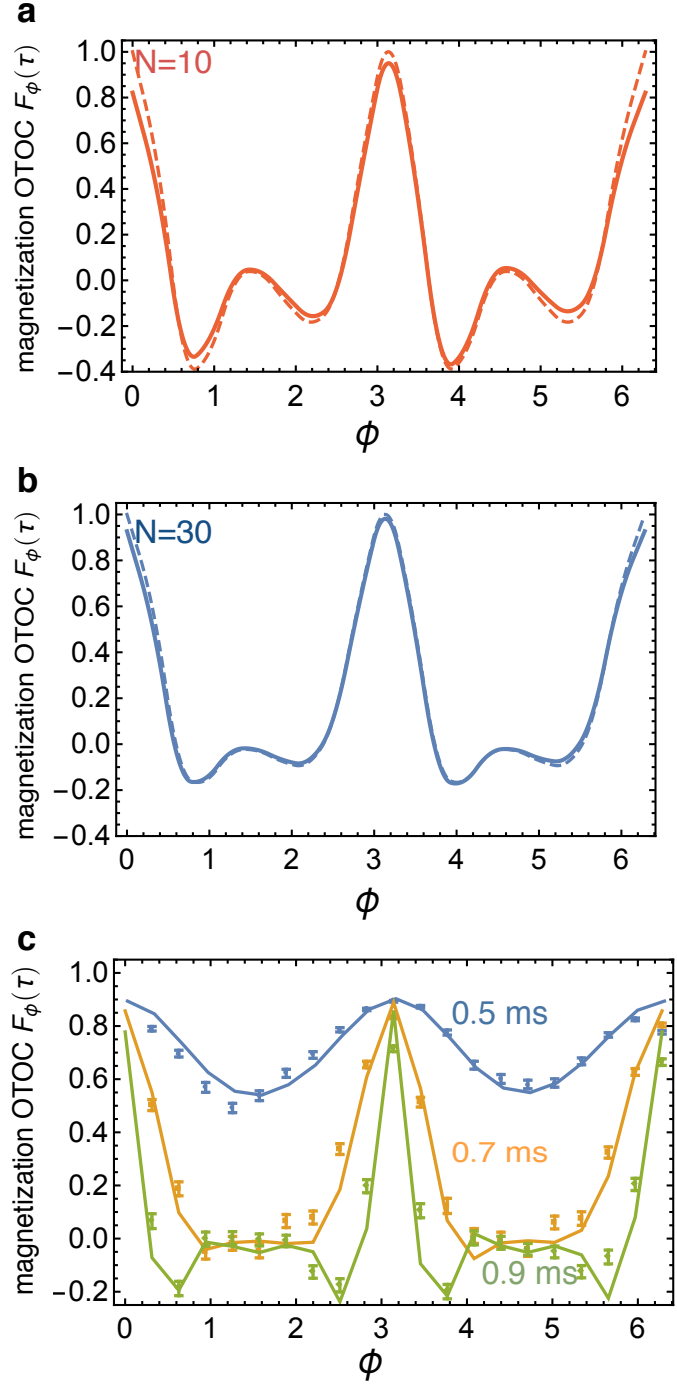}
    \caption{\textbf{Effect of spin-motion coupling on magnetization OTOC $F_\phi(\tau)$} \textbf{a, b,} $F_\phi(\tau)=\left(2/N\right)\langle \hat S_x\rangle$ at $\tau=0.9\,$ms using $J/\hbar = 4374$~s$^{-1}$ for $N=10$ and $N=30$, respectively. In each panel the solid line includes the COM mode fluctuations and magnetic field noise. The effect of COM mode fluctuations is less visible in larger crystals. \textbf{c,} Experiment-theory comparisons for $N=48$ using the same parameters as Fig.~3a.}
    \label{fig:Sx}
\end{figure}
The COM fluctuations manifest as imperfect knowledge of $\omega_z$ in each run of the experiment. This means that for each experimental run the actual COM frequency is in fact $\omega_{\rm COM}=\omega_z + \delta \omega_z$, resulting in an actual detuning $\delta_{\rm COM}= \delta- \delta \omega_z$ for the first arm of the experiment and $\delta_{\rm COM}= -\delta - \delta \omega_z$ in the reversal. In light of the imperfection in the reversal we label the argument of the displacement operator for the first (second) arm of the experiment with $\alpha$ ($\beta$), with $\hat{D}(\alpha)= \exp\left[\alpha \hat{a}_0^\dagger-\bar \alpha \hat{a}_0 \right]$. Additionally we use $J=J(\tau,0)$ and $\tilde J=J(2\tau, \tau)$ to denote the spin-spin interactions in the two experimental arms. Note that we use the full expressions for $\alpha$ and $J$ given by Eqs.~\eqref{eq:alphafull} and \eqref{eq:Jijfull}.  

In principle, the ODF phase in Eq.~\eqref{eq:Hsupp} can be different in the two arms. In the following, we model the experimental protocol where 
these phases, denoted by $\varphi_1$ and $\varphi_2$, are chosen to satisfy $\varphi_2-\varphi_1=2 \pi (t_\pi/\tau)$ with $t_\pi=75$~$\mu$s the length of the $\pi$-pulse. 

Here, we outline the procedure used to quantify the effect of center of mass mode fluctuations, modeled as a Gaussian noise with width $ \sigma/2\pi =\Delta_{\rm COM}=125(50)$~Hz on the fidelity $\mathcal{F}_\phi (\tau)$ and spin component $\langle \hat S_x \rangle$. 

In order to facilitate experimental comparisons we can easily include the effects of a static magnetic field noise within the formalism presented above. To this end we modify $\hat{\mathcal{U}}_{\rm SP}$ by adding the term $B\sum_j \hat \sigma_j^z$, where $B$ is sampled from a Gaussian distribution with $ \sigma_B/2\pi= \Delta_{\rm B} \sim 40$~Hz. This is equivalent to the substitutions $\alpha \to \alpha + B \tau$ and $\beta \to \beta + B \tau$ in the following expressions.

\subsection{Fidelity $\mathcal{F}_\phi(\tau)$.}
We take advantage of the all-to-all nature of the interactions and use the manifold of collective Dicke states to study the unitary evolution of the system. We find that the thermally averaged fidelity is given by
% \begin{widetext}
\begin{equation}
\label{eq:F_phonons}
\begin{aligned}
 \langle \mathcal{F}_\phi(\tau)\rangle_{\rm th} &= \sum_{m, m^\prime, l, l^\prime=0 }^N \frac{\sqrt{\binom{N}{m} \binom{N}{m^\prime}}  \sqrt{\binom{N}{l} \binom{N}{l^\prime}}} {2^{2N}} {d}_{M^\prime/2,M/2}^{N/2}(\phi)\bar{d}_{L^\prime/2,L/2}^{N/2}(\phi)e^{-i( J M^2+\tilde J M^{\prime^2})\tau/2} e^{i( J L^2+\tilde J L^{\prime^2})\tau/2} \\
&\times e^{i (\theta_{L,L^\prime}+\theta_{M,M^\prime})}\exp\left[-(\bar n+1/2) (\vert \gamma \vert^2+ \vert \tilde \gamma \vert^2 )\right] \mathcal{I}_0\left(2 e^{\beta_{\rm th}/2}\vert \gamma\vert \vert \tilde \gamma \vert \bar n \right)\, ,
\end{aligned}
\end{equation}
% \end{widetext}
where the capitalized form of the summation indices corresponds to $A=N-2a$ for $A=M, M', L, L'$, and $a=m,m',l,l'$. Here $\gamma\equiv M \alpha+M'\beta$, $\tilde \gamma\equiv L \alpha+L'\beta$, $\theta_{M,M'}\equiv{\rm Im}\left[M'M \bar \beta\alpha\right]$, and $\theta_{L,L'}\equiv{\rm Im}\left[L'L \bar \beta \alpha\right]$. We have used $d_{M^\prime/2,M/2}^{N/2}(\phi)$ and $d_{L/2,L^\prime/2}^{N/2}(\phi)$ to denote the Wigner $d$ matrices in the $z-x-z$ convention, which are related to the $z-y-z$ convention matrices by a multiplicative factor,  $d_{M^\prime/2,M/2}^{N/2}(\phi)= d_{M^\prime/2,M/2}^{N/2\, (z-y-z)}(\phi)i^{(M/2- M'/2)}$. Here $\bar{n}$ is the thermally averaged mode quanta and $\beta_{\rm th}=1/k_B T$. Finally $\mathcal{I}_0$ is the modified Bessel function with power series expansion $\mathcal{I}_0(x)=1+x^2/4+\dots$. 

We note that the COM fluctuations lead to a strong decay of the fidelity signal at $\phi=\pi$. This is because for the experimentally implemented protocol the spin-dependent force has approximately the same phase in both arms of the sequence and the resulting spin-dependent displacements in the two arms are approximately opposite, $\alpha \approx -\beta$. Hence, a $\phi=\pi$ rotation nearly doubles the spin-dependent displacement, leading to spin-phonon entanglement, at the end of the sequence.  This is in contrast to the static magnetic field noise, as the latter is completely eliminated by the $\phi = \pi$ (echo) sequence. The reverse is true for $\phi = 0, 2\pi$, where the effect of COM fluctuations are suppressed and static magnetic field noise leads to a decay of the fidelity signal.

Finally, as discussed in section~\ref{sec:master} the experiment operates in the regime where $\Gamma_{\rm el}\gg \Gamma_{\rm ud},\Gamma_{\rm du}$. In this regime, and for experimentally relevant times, the decay of fidelity can be approximated by $\mathcal F_\phi(\tau)\to e^{-N \Gamma \tau}  \left[\mathcal F_\phi(\tau)\right]_{\Gamma\to 0}$, where $\Gamma= (\Gamma_{\rm el}+ \Gamma_{\rm ud} + \Gamma_{\rm du})/2$. We have used this approximation and the expression given in Eq.~\eqref{eq:F_phonons} for the theoretical predictions in Fig.~3(a) in the main text. In Fig.~3(b) the expression was modified to include the $\pi$-pulses (echo) in the middle of each arm of the MQC sequence. 

\subsection{Magnetization $\langle \hat S_x \rangle$.}
Using the same formalism we can find an analytic expression for the spin component $\langle \hat S_x \rangle$
% \begin{widetext}
\begin{equation}
\begin{aligned}
\biggl\langle\hat S_x\biggr \rangle_{\rm th}=\mathrm{Re} \biggl[ &\sum_{m, m^\prime, p^\prime} \frac{\sqrt{\binom{N}{m} \binom{N}{m'} }\sqrt{p'(N/2+p'+1)}}{2^N} e^{-i J (M^2-M^{\prime^2}) \tau/2}e^{-i \tilde J ( (P^\prime-2)^2-P^{\prime^2}) \tau/2}\\
&d_{N/2-m^\prime, N/2-p^\prime}^{N/2}(-\phi) d_{N/2-(p^\prime -1), N/2-m}^{N/2}(\phi) e^{i (\nu_1-\nu_2)} e^{-\vert\gamma\vert^2(\bar n +1/2)} \biggr ]\, ,
\end{aligned}
\label{eq:Sx}
\end{equation}
% \end{widetext}
where $P^\prime=N-p^\prime$, $\nu_1\equiv (M^\prime P^\prime - M (P^\prime-2)) {\rm Im} \left[ \bar \alpha \beta \right]$, $\nu_2\equiv{\rm Im} \left[(M^\prime \alpha+P^\prime \beta)((P'-2) \bar\beta+M \bar \alpha\right)$, and $\gamma\equiv \left(M-M^\prime\right) \alpha - 2\beta$.

Evaluating the expression for $\langle \hat S_x\rangle$ given by Eq.~\eqref{eq:Sx} in the presence of static magnetic field noise and for large number of spins is computationally expensive. However we observe that the effect of COM frequency fluctuations on $\langle \hat S_x \rangle$ decreases as the size of the system $N$ increases. This is illustrated in Fig.~\ref{fig:Sx}(a) and (b), where in both panels the solid lines include the decoherence effects due to COM fluctuations, as well as the static magnetic field noise, and the dashed lines are in the absence of these two decoherence effects. In Fig.~\ref{fig:Sx}(c) we show the experimental data for $\langle\hat S_x \rangle/(N/2)$ corresponding to the fidelity data shown in Fig.~3a. As discussed in the previous section the experiment operates in the regime where $\Gamma_{\rm el}\gg \Gamma_{\rm ud}, \Gamma_{\rm du}$. In this regime, and for experimentally relevant times, the decay of $\langle \hat S_x \rangle$ can be approximated by a multiplicative factor $e^{-\Gamma \tau}$. Here we use $\Gamma=62$~$s^{-1}$.

\twocolumngrid

\bibliographystyle{apsrev4-1}
\bibliography{main}

\noindent\textbf{Acknowledgements}\\
We thank Philipp Hauke, John Price, and Shimon Kolkowitz for discussions and careful reading of our manuscript and gratefully acknowledge Joseph Britton and Brian Sawyer for preceding experimental contributions to this work. 
Supported by Defense Advanced Research Projects Agency (DARPA), NSF grant PHY 1521080, JILA-NSF grant PFC-1125844, the Army Research Office (ARO), and the Air Force Office of Scientific Research and its Multidisciplinary University Research Initiative (AFOSR-MURI) (A.M.R.) and by a National Research Council Research Associateship Award at NIST (J.G.B. and M.L.W.).
All authors acknowledge financial support from NIST.

\noindent\textbf{Author contributions}\\
J.G.B and J.B. conducted the experiment. The theoretical modeling was done by M.G., A.S-N., M.L.W., and A.M.R. All authors jointly interpreted and discussed the experimental data.

\end{document}